\documentclass[11pt,a4paper]{article}

\pdfoutput=1

\usepackage{jheppub}
\usepackage{latexsym}
\usepackage{multirow}
\usepackage{color}
\usepackage[usenames,dvipsnames,table]{xcolor}
\usepackage{graphicx}
\usepackage{epsfig} 
\usepackage{epsf}   
\usepackage{dcolumn}
\usepackage{bm}
\usepackage{dcolumn}
\usepackage{textcomp}
\usepackage{float}
\usepackage{subfig}
\usepackage{hypcap}
\usepackage[]{hyperref}
\usepackage{makecell}
\usepackage{color}
\usepackage{pifont}
\usepackage{appendix}
\usepackage{amsmath}
\usepackage{multirow,bigdelim}  
\usepackage{lineno}
\usepackage[normalem]{ulem}
\graphicspath{{fig/}}
\usepackage{physics}
\usepackage{simpler-wick}
\usepackage{slashed}

\def\be{\begin{equation}}
\def\ee{\end{equation}}
\definecolor{maroon}{rgb}{0.5, 0.0, 0.0}	
\definecolor{arsenic}{rgb}{0.23, 0.27, 0.29}

\newcommand{\comment}[1]{}
\usepackage{amssymb}
\hypersetup{
	bookmarks=true,         
	unicode=false,          
	pdftoolbar=true,        
	pdfmenubar=true,        
	pdffitwindow=true,   
	pdfstartview={FitH},    
	pdfsubject={dark NSI},   
	pdfnewwindow=true,      
	pdfcreator={RevTeX},
	colorlinks=true,     
	linkcolor=Blue,         
	citecolor=Blue,        
	filecolor=Blue,     
	urlcolor=Blue,           
}

\preprint{}

\title{\boldmath Dark NSI \& neutrino oscillations : probing via $\delta_{CP}$ measurements at DUNE and T2HK}

\author[a]{Dharitree Bezboruah,} 
\author[b]{Abinash Medhi,}
\author[a]{Moon Moon Devi,}

\affiliation[a]{Department of Physics, Tezpur University, Napaam, Sonitpur, Assam-784028, India}
\affiliation[b]{Department of Physics, G.L. Choudhury College, Barpeta Road, Assam 781315, India}

\emailAdd{dbbphy1@tezu.ernet.in}
\emailAdd{abinashmedhi0@gmail.com}
\emailAdd{devimm@tezu.ernet.in}

\date{\today}

\abstract{ We investigate the possibility of neutrinos interacting with a scalar dark matter field and the resulting implications for neutrino oscillations in the long-baseline sector. As our Universe is predominantly composed of dark matter, neutrinos propagating over astrophysical and terrestrial baselines inevitably traverse a dark matter background. The coherent forward scattering of neutrinos in such a background induces a medium-dependent correction to the mass-squared term in the effective neutrino Hamiltonian having opposing signs for neutrinos and antineutrinos. We study how the elements of this correction matrix, arising from coherent forward scattering of neutrinos with scalar dark matter background referred to as dark non-standard interactions (dark NSI), modify neutrino oscillation probabilities. Furthermore, we also study the effect of the off-diagonal elements and the associated phases on the measurement of leptonic CP violating phase focusing on the upcoming long-baseline superbeam experiments DUNE and T2HK. We show that dark NSI can lead to substantial enhancement or suppression of CP-violation sensitivity, depending on the true values of the dark NSI phases $\phi_{\alpha \beta}$. We further explored how the synergy of DUNE and T2HK can effectively mitigate the degeneracies due to the dark NSI phases, and can restore or even enhance the CP sensitivity as compared to the standard oscillation scenario.}


\begin{document}
\maketitle


\section{Introduction}

Although the Standard Model (SM) of particle physics has been exceptionally successful in describing the fundamental constituents of matter and their interactions, it remains an incomplete framework that fails to address several key mysteries. While it provides a unified description of electromagnetic, weak, and strong interactions, the SM lacks a natural mechanism to explain the nonzero masses and flavor mixing of neutrinos. The discovery of neutrino oscillations by the Super-Kamiokande (SK)~\cite{Super-Kamiokande:1998kpq} and Sudbury Neutrino Observatory (SNO)~\cite{SNO:2002tuh} served as a landmark in particle physics, offering the first experimental proof that neutrinos possess mass and they oscillate while traversing through space. Various neutrino experiments continue to explore the parameters governing oscillations~\cite{Super-Kamiokande:2004orf,KamLAND:2004mhv,MINOS:2008kxu,MINOS:2011neo}, further refining our understanding of these fundamental particles. Currently, the major unresolved questions in the three neutrino paradigm are the mass ordering (MO), the octant of $\theta_{23}$, the value of the leptonic CP phase $\delta_{CP}$, and the values of the absolute neutrino masses. With the advent of future high-precision experiments, it is a critical time to examine whether experimental data are converging on the standard three-flavor framework or if hints of new physics are emerging. 

One compelling avenue for new physics is the presence of Non-Standard Interactions (NSI)~\cite{Liao:2016orc,Friedland:2012tq,Coelho:2012bp,Rahman:2015vqa,Coloma:2015kiu,deGouvea:2015ndi,Liao:2016hsa,Forero:2016cmb,Huitu:2016bmb,Bakhti:2016prn,Kumar:2021lrn,Agarwalla:2015cta,Agarwalla:2014bsa,Agarwalla:2012wf,Blennow:2016etl,Blennow:2015nxa,Deepthi:2016erc,Masud:2021ves,Soumya:2019kto,Masud:2018pig,Masud:2017kdi,Masud:2015xva,Ge:2016dlx,Fukasawa:2016lew,Chatterjee:2021wac,Chaves:2021kxe,Brahma:2023wlf,Davoudiasl:2023uiq,Chatterjee:2020kkm,Choubey:2014iia,Singha:2021jkn,Denton:2018xmq,Denton:2020uda,Farzan:2015hkd, Ge:2018uhz,Medhi:2021wxj, Medhi:2023ebi,Medhi:2022qmu,Babu:2019iml,Bezboruah:2024yhk,Sarker:2024ytu} within the neutrino sector. NSI refers to interactions between neutrinos and other particles that are not formulated in the SM of particle physics. These interactions could arise from physics at a higher energy scale and often manifest as effective four-fermion interactions mediated by heavy particles. Such interactions have been extensively studied in the context of neutrino oscillations, scattering experiments, and astrophysical phenomena. However, an alternative and intriguing possibility is the existence of interaction of neutrinos with dark matter (DM) particles, termed the dark NSI~\cite{Ge:2019tdi}. Several cosmological and astronomical sources have established that DM is a constituent entity of the universe \cite{Rubin:1980zd,Zwicky:1933gu,Clowe:2006eq,Planck:2013pxb}. However, the nature and properties of DM are still a pressing question for particle physicists. Considering the current literature, the mass of DM can range over many orders of magnitude, ranging from fuzzy DM to primordial black holes \cite{Ferreira:2020fam}, and there are a wide variety of particles which can act as viable DM candidates \cite{Bergstrom:2009ib}. We are particularly interested in ultra-light DM particles, which are bound to be bosonic in nature as restricted by quantum statistics. The bosonic DM particles can be either be a scalar or a vector. In this work, we are focusing on the interaction of neutrinos with ultra-light scalar DM particles in a model-independent way.

There has been a vast range of literature exploring how interaction of neutrino with DM can imprint observable signatures in oscillation experiments. A key focus has been on the effects of ultra-light DM fields. For instance, DM couplings with neutrinos can induce periodic variations in neutrino masses and mixing parameters \cite{Berlin:2016woy}, an effect that is strongly constrained by the absence of time-modulated solar neutrino fluxes. When averaging over these fast oscillations, the resulting phenomena can manifest as Distorted Neutrino Oscillations (DiNOs), offering a new target for future long-baseline and reactor experiments like DUNE and JUNO \cite{Krnjaic:2017zlz}. The framework has also been extended to include fuzzy DM, where coherent forward scattering between neutrinos and the DM field can induce a noticeable shift in the oscillation probabilities \cite{Brdar:2017kbt}. DM neutrino interactions can also be modeled by considering the DM background as an effective medium that modifies the neutrino Hamiltonian. This generalization applies to various DM Lorentz structures, including scalar, vector, and tensor backgrounds \cite{Capozzi:2017auw}. Furthermore, specific DM models, such as asymmetric DM captured in the Sun, have been shown to potentially induce a Dark MSW effect \cite{Capozzi:2018bps}, thereby altering the observable solar neutrino spectrum. The theoretical implications also extend to extreme astrophysical environments; the self-energies of scalar and fermionic DM environments are crucial for understanding neutrino propagation in contexts like supernovae and the early universe \cite{Nieves:2018vxl}. Finally, the study of neutrinophilic axion-like dark matter (ALP) has revealed that derivative ALP and neutrino couplings can significantly affect neutrino oscillations and their propagation in various astrophysical settings \cite{Huang:2018cwo}.

In this work, we concentrate on neutrino interactions with a complex scalar dark matter field, giving rise to dark NSI as discussed in ref.~\cite{Ge:2019tdi}. Unlike conventional NSI, which typically involves contact interactions that modify neutrino scattering processes, the presence of a coherent DM background induces a correction to the neutrino mass-squared matrix. The correction term has opposing signs for neutrinos and antineutrinos. Such corrections can significantly alter oscillation probabilities and introduce new sources of CP-violating effects. We investigate the phenomenological implications of dark NSI for future long-baseline neutrino experiments, with particular emphasis on DUNE \cite{DUNE:2020ypp} and T2HK \cite{Hyper-KamiokandeProto-:2015xww}. Our analysis focuses on how the moduli and phases of dark NSI parameters affect neutrino oscillation probabilities and the measurement of the leptonic CP-violating phase. By exploring the complementarity between DUNE and T2HK, we assess the extent to which experimental synergy can mitigate parameter degeneracies and enhance sensitivity to CP violation in the presence of dark NSI. Our study provides new insights into the interplay between neutrino oscillation physics and the dark sector, highlighting the potential of next-generation experiments to probe interaction of neutrinos with dark matter.

This paper is organized in the following way. In section~\ref {sec:formalism}, we present the detailed formalism of dark NSI. In section~\ref{sec:methodology}, experimental configurations of the LBL experiments and the simulation framework are specified. In section~\ref{sec:result}, we discuss the findings of our study, and in section~\ref{sec:summary}, we provide a summary of our work.

\section{Theoretical formalism}\label{sec:formalism}

When neutrinos propagate through matter, the coherent forward scattering with environmental fermions introduces an effective matter potential \cite{Wolfenstein:1977ue,Mikheyev:1985zog} and modifies the vacuum Hamiltonian. This propagation Hamiltonian is further modified due to subdominant effects, such as non-standard interactions \cite{Liao:2016orc,Friedland:2012tq,Coelho:2012bp,Rahman:2015vqa,Coloma:2015kiu,deGouvea:2015ndi,Liao:2016hsa,Forero:2016cmb,Huitu:2016bmb,Bakhti:2016prn,Kumar:2021lrn,Agarwalla:2015cta,Agarwalla:2014bsa,Agarwalla:2012wf,Blennow:2016etl,Blennow:2015nxa,Deepthi:2016erc,Masud:2021ves,Soumya:2019kto,Masud:2018pig,Masud:2017kdi,Masud:2015xva,Ge:2016dlx,Fukasawa:2016lew,Chatterjee:2021wac,Medhi:2023ebi,Chaves:2021kxe,Brahma:2023wlf,Davoudiasl:2023uiq,Chatterjee:2020kkm,Choubey:2014iia,Singha:2021jkn,Denton:2018xmq,Denton:2020uda,Farzan:2015hkd,Ge:2018uhz, Medhi:2021wxj}, Lorentz Invariance Violation \cite{Kostelecky:2003cr, SNO:2018mge,Mewes:2019dhj, Huang:2019etr, ARIAS2007401,LSND:2005oop,MINOS:2008fnv,MINOS:2010kat,IceCube:2010fyu,MiniBooNE:2011pix,DoubleChooz:2012eiq, Sarker:2023mlz,Sarkar:2022ujy,Majhi:2022fed}, decoherence \cite{PhysRevD.56.6648,Benatti:2000ph,PhysRevD.100.055023,PhysRevD.95.113005,Lisi:2000zt}, etc. Several medium-dependent effects have also been studied recently, considering the propagation of neutrinos in a medium of dark matter (DM) \cite{Berlin:2016woy, Berlin:2016bdv,Capozzi:2017auw,Krnjaic:2017zlz,Brdar:2017kbt,Liao:2018byh,Capozzi:2018bps,Nieves:2018vxl,Huang:2018cwo,Pandey:2018wvh,Nieves:2018ewk}, and dark energy \cite{Gu:2005eq,Ando:2009ts,Ciuffoli:2011ji,Klop:2017dim}. It is evident from astrophysical and cosmological studies that the local DM energy density is $\rho_\chi \equiv 0.47~\rm{GeV/cm^3}$ and the number density of the constituent dark matter particles $(n_\chi)$ is inversely proportional to their mass $(m_\chi)$ \cite{Ge:2019tdi}. This implies a significant abundance of light DM. It may be noted that, if dark matter particles possess masses below $\sim$100 eV, quantum statistics constrain them to be bosonic in nature; either a scalar or a vector boson. In this work, we consider the propagation of neutrinos in a medium full of ultralight scalar DM $\phi$, with a number density $n_\phi$. The interaction between $\phi$ and the neutrinos can be represented by the following Lagrangian

\be
-\mathcal{L}=\frac12 m_\phi^2 \phi^2 +\frac12 M_{\alpha \beta} \bar \nu_\alpha \nu_\beta + y_{\alpha \beta}\phi \bar \nu_{\alpha} \nu_\beta+ \rm{h.c.}.
\ee

\noindent Here, $y_{\alpha \beta}$ denotes the Yukawa coupling between the neutrino flavors and the scalar DM field $\phi$. The masses of the DM field and the neutrino mass matrix are denoted by $m_\phi$ and $M_{\alpha \beta}$, respectively. The corresponding correction to the Dirac equation is as follows,

\begin{figure}[h!]
    \centering
    \includegraphics[width=0.36\linewidth]{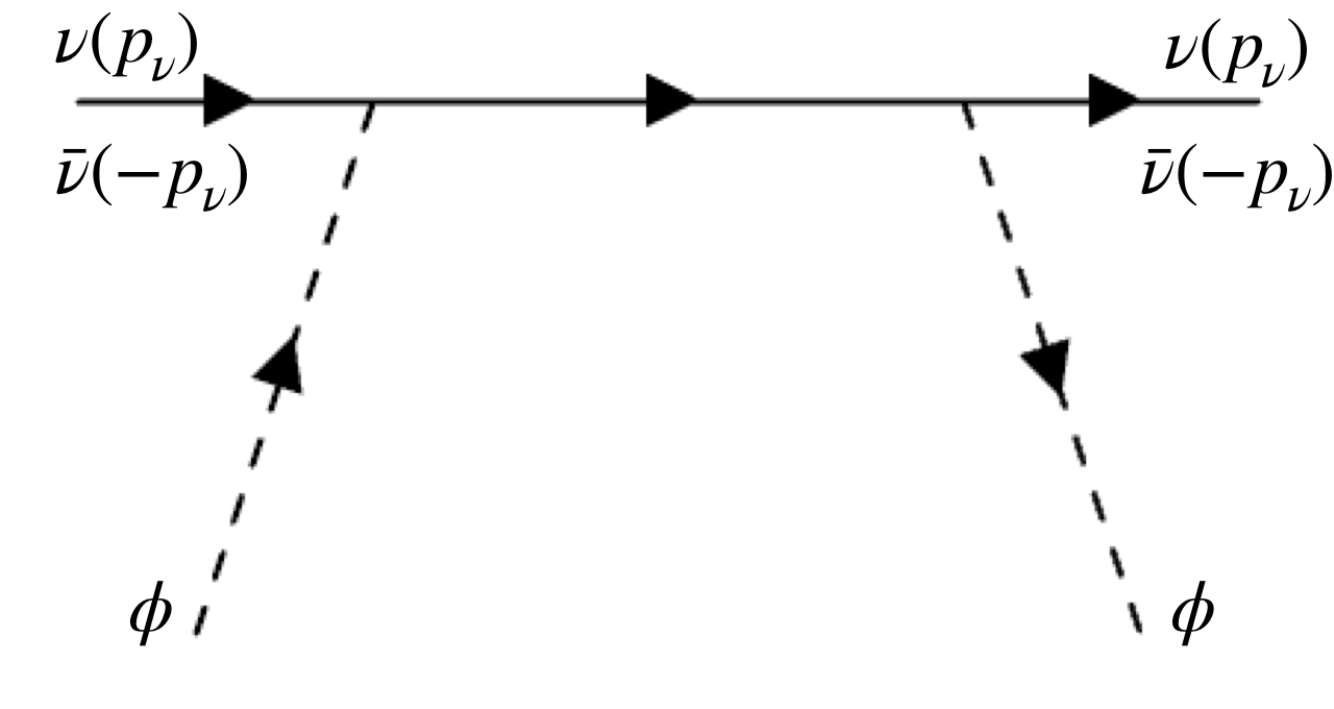}\quad
    \includegraphics[width=0.4\linewidth]{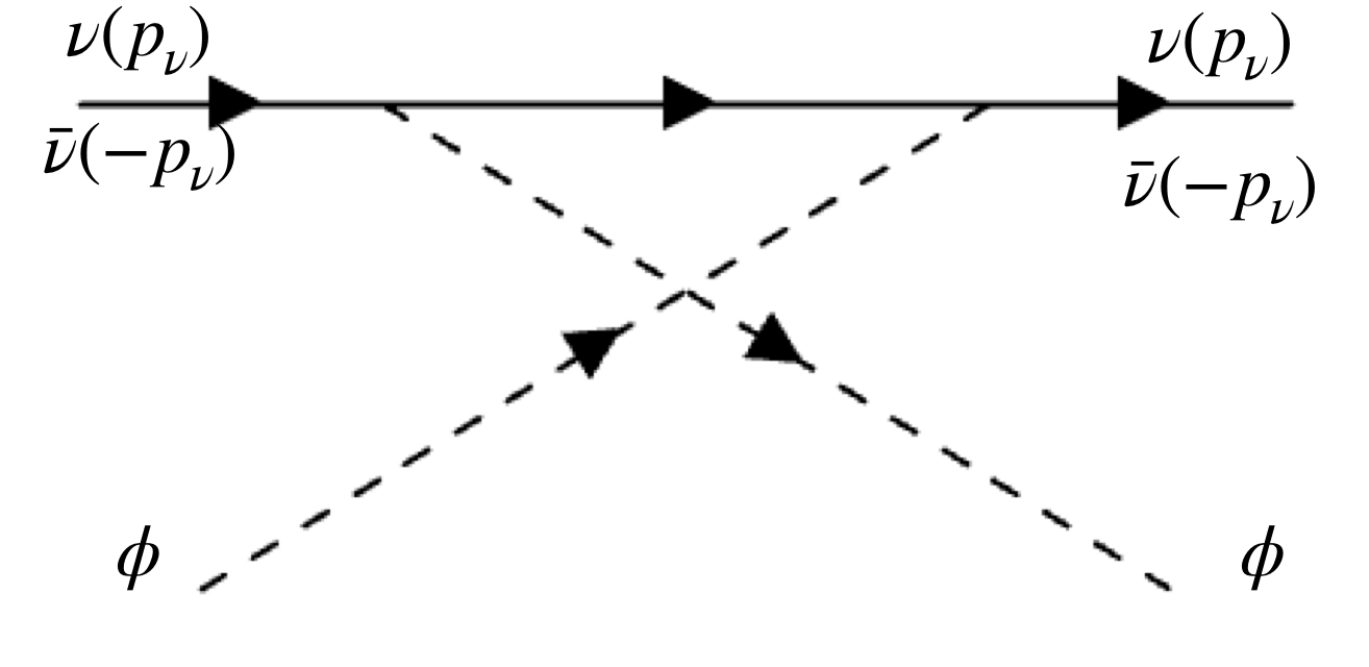}
    \caption{Feynman diagram for neutrino forward scattering with scalar $\phi$ \cite{Ge:2019tdi}.}
    \label{fig:feyn}
\end{figure}

\be
i(\slashed{\partial} - M_{\alpha \beta} - y_{\alpha \beta} \phi ) \nu_\beta =0.
\ee

For the propagation of neutrinos in a medium of DM $\phi$, both $\phi$ and neutrinos will be present as real particles. For coherent forward scattering with zero momentum transfer, the initial and final states of the neutrino and DM remain identical ($p_\nu = p_\nu'$ and $p_\phi = p_\phi'$). This interaction can be described by the transition matrix element,

\begin{equation}
\langle \nu_\alpha(p_\nu) \phi(p_\phi) | T \exp \left( i \int d^4x \mathcal{L} \right) | \nu_\beta(p_\nu) \phi(p_\phi) \rangle.
\end{equation}

The leading-order contribution arises from the second-order $S$-matrix element involving two $\phi \bar{\nu} \nu$ vertices. Using second quantization, we can expand the neutrino field as $\nu = au + b^\dagger v$. As a consequence, only $\nu$ can contract with $\ket{\nu}$ and $\bar{\nu}$ can contract with $\bra{\nu}$. By applying Wick's theorem, we identify the non-vanishing contractions contributing to the forward scattering amplitude for the $\nu_\alpha \to \nu_\beta$ transition. As shown in eq.~\eqref{eq:wick_nu}, the first term corresponds to the $s$-channel exchange, whereas the second term represents the $u$-channel contribution, with the internal lines denoting the respective propagators for the scalar and neutrino fields.

\begin{equation}\label{eq:wick_nu}
\mathcal{S}_{\alpha \beta} \sim 
\frac 12 \langle
 \wick{
        \c1{\nu_\beta} \c2 \phi
        \vert
        (-iy_{ij} \overline{\c1 \nu}_i \c2\phi \c3 \nu_j) ( iy_{kl}^* \overline{\c3 \nu}_k \c4 \phi^*  \c5 \nu_l 
        \vert
        \c5 \nu_\alpha \c4 \phi
  }
\rangle +
\frac 12 \langle
 \wick{
        \c1{\nu_\beta} \c2 \phi
        \vert
        (-iy_{ij} \overline{\c1 \nu}_i \c4\phi \c3 \nu_j) ( iy_{kl}^* \overline{\c3 \nu}_k \c2 \phi^*  \c5 \nu_l 
        \vert
        \c5 \nu_\alpha \c4 \phi
  }
\rangle.
\end{equation}

Similarly, for the $\bar{\nu}_\alpha \to \bar{\nu}_\beta$ transition, the second order term in the $S$-matrix has the following form,

\begin{equation}\label{eq:wick_barnu}
\mathcal{\bar S}_{\alpha \beta} \sim 
\frac 12 \langle
 \wick{
        \c1{\bar \nu_\beta} \c2 \phi
        \vert
        (-iy_{ij} \c5 \nu_i \c2\phi \c3 \nu_j) ( iy_{kl}^* \overline{\c3 \nu}_k \c4 \phi^*  \c1 \nu_l 
        \vert
        \overline{\c5 {\nu}}_\alpha \c4 \phi
  }
\rangle +
\frac 12 \langle
 \wick{
        \c1{\bar \nu_\beta} \c2 \phi
        \vert
        (-iy_{ij} \c5 \nu_i \c4\phi \c3 \nu_j) ( iy_{kl}^* \overline{\c3 \nu}_k \c2 \phi^*  \c1 \nu_l 
        \vert
         \overline{\c5 {\nu}}_\alpha \c4 \phi
  }
\rangle.
\end{equation}

It should be noted here that the permutation of fermion operators introduces a fundamental difference between neutrino and antineutrino transitions, 

\begin{itemize}
    \item In case of the $\nu_\alpha \to \nu_\beta$ transition, field operators $\nu$ and $\bar{\nu}$ contract directly with the external states $|\nu_\alpha\rangle$ and $\langle \nu_\beta|$. No additional permutations are required, leading to the relation $\delta S_{\beta\alpha} = \bar{u}_\beta \delta \Gamma_{\beta\alpha} u_\alpha$.
    
    \item Whereas for the $\bar{\nu}_\alpha \to \bar{\nu}_\beta$ transition, pairing the field operators with the antineutrino states $|\bar{\nu}\rangle = b^\dagger |0\rangle$ requires an odd number of permutations. This introduces a factor of $-1$ due to the anti-commutation relations, such that $\delta \bar{S}_{\beta\alpha} = -\bar{v}_\alpha \delta \Gamma_{\alpha\beta} v_\beta$.
\end{itemize}

Following the Feynman rules for the diagrams in figure~\ref{fig:feyn}, the two-point correlation function $\delta \Gamma_{\alpha \beta}$ is obtained as,

\begin{eqnarray} \label{eq:Lamba}
\delta \Gamma_{\alpha \beta} &=& \frac{ i 2 \rho_x}{m_\phi^2} \sum_k y_{\beta k} y^*_{k \alpha} \left[\frac{(\slashed{p}_\nu+ \slashed{p}_\phi)+m_\nu}{(p_\nu + p_\phi)^2 - m
_\nu^2 } + \frac{(\slashed{p}_\nu- \slashed{p}_\phi)+m_\nu}{(p_\nu - p_\phi)^2 - m_\nu^2} \right].
\end{eqnarray}

Here, $\rho_x$ is the density of the scalar DM $\phi$. For on-shell neutrinos, the denominators simplify to $(p_\nu \pm p_\phi)^2 - m_\nu^2 = p_\phi^2 \pm 2 p_\nu \cdot p_\phi$. In the limit of non-relativistic DM ($p_\phi \ll p_\nu$), where $p_\phi \approx (m_\phi, 0)$, the scalar product reduces to $p_\nu \cdot p_\phi \approx E_\nu m_\phi$. Substituting these into eq. \eqref{eq:Lamba} we get

\begin{equation}
\delta \Gamma_{\alpha \beta} = \frac{ i 2 \rho_x}{m_\phi^2} \sum_k y_{\beta k} y^*_{k \alpha} \left[\frac{ \slashed{p}_\phi}{2 m_\phi E_\nu} \right ].
\end{equation}

As the DM around Earth is non-relativistic, we need to consider only
the dominating temporal component, $\slashed{p}_\phi \approx m_\phi \gamma^0$. Therefore, the resulting leading-order correction due to dark NSI is

\begin{equation}\delta \Gamma_{\alpha \beta} \approx \frac{\rho_\chi}{m_\phi^2 E_\nu} \sum_k y_{\beta k} y^*_{k \alpha} \gamma^0.
\end{equation} 

As we have mentioned earlier, when transitioning from neutrino to antineutrino mode, the momentum in the propagator receives a minus sign
to account for the opposite fermion flow. Therefore, the Hamiltonian in the presence of dark NSI can be written as,

\be \label{eq:dark_Hamiltonian}
 H_{DNSI}= \frac{M^2 \pm \delta M^2}{2 E_\nu} \pm V_{SI},
 \ee
 
\noindent where $\delta M_{\alpha \beta}^2 \equiv \pm \frac{2\rho_x}{m_\phi^2}\sum_{j}y_{\alpha j} y_{j \beta}^*$. The effective potential induced by dark NSI has a distinct $E_\nu^{-1}$ energy dependence. Due to this energy dependence, the dark NSI contribution can be physically interpreted as a correction to the neutrino mass-squared term in the neutrino Hamiltonian rather than a traditional matter potential. This yields unique phenomenological consequences, as the dark matter density $\rho_\phi$ effectively modifies the mass-squared differences governing neutrino oscillations.

In order to explore the effect of dark NSI in neutrino oscillation experiments, we have used a similar parametrization $\delta M^2$ as that of scalar NSI \cite{Ge:2018uhz, Medhi:2021wxj} but with $S_{m}^2$ as the characteristic scale \cite{Ge:2019tdi}, 

\be
\delta M^2= S_{m}^2\begin{pmatrix}
     d_{ee}& d_{e \mu} & d_{e \tau}\\
    d^*_{e\mu}& d_{\mu \mu} & d_{\mu \tau}\\
     d^*_{e\tau}& d^{*}_{ \mu \tau} & d_{\tau \tau}
     \end{pmatrix}.
\ee
     The $d_{\alpha \beta}$ are dimensionless parameters that signify the strength of dark NSI. Here, $S_{m}^2$ is introduced as a scaling term with the dimension of neutrino mass-squared. In this work we have used 
     
    \begin{equation}
        S_{m}^2 = 2.53 \times 10^{-3} eV^2,
    \end{equation}
    which corresponds to a characteristic atmospheric mass squared difference. To preserve the hermiticity of the Hamiltonian, the diagonal parameters are considered to be real, and the off-diagonal parameters are considered to be complex ($d_{\alpha \beta} = | d_{\alpha \beta} | e^{i \phi_{\alpha \beta}}$ ).

The correction due to dark NSI scales as $\delta M_{\alpha \beta}^2 \propto y_{\alpha \beta}^2/m_\phi^2$. This encompasses a large parameter space because the constraints on the coupling $y$ and the mediator mass $m_\phi$ are relatively weak in the ultralight regime. Neutrino oscillation experiments are best suited to explore this sector, as they are sensitive to coherent forward scattering with zero momentum transfer. In contrast, other kinematic processes involve significant momentum transfer, where the interaction strength scales as $\propto y^2/(q^2-m_\phi^2)$. In the limit of an ultralight mediator ($|q^2| >> m_\phi^2$), the effect is suppressed by the momentum exchange rather than the mediator mass. Consequently, the signal in non-oscillation searches is naturally attenuated since $y^2/q^2 << y^2/m_\phi^2$.  Therefore, neutrino oscillation is most favourable to probe the regions with proportionally tiny mass and coupling that is formidable for any other experiments to reach.

\section{Experiment and simulation details}\label{sec:methodology}

In this work, we investigate the impact of dark NSI on the measurement of the CP-violating phase at upcoming LBL experiments. In this section, we provide the technical specifications for DUNE \cite{DUNE:2020ypp} and T2HK \cite{Hyper-KamiokandeProto-:2015xww}, followed by a detailed description of our simulation framework.

\subsection{DUNE} The Deep Underground Neutrino Experiment (DUNE) \cite{DUNE:2020ypp} is a future flagship LBL accelerator experiment hosted by Fermilab, with a baseline of 1300 km that will run from Fermilab to Stanford Underground Research Facility (SURF) in South Dakota. Neutrinos will be produced by the Long Baseline Neutrino Facility (LBNF) from a 120 GeV proton beam having 1.2 MW power with a capacity to deliver 1.1 $\times$ $10^{21}$ proton on target (POT) per year. DUNE will utilize an intense, on-axis, broad-band muon neutrino beam, peaked at 2.5 GeV for optimal oscillation sensitivity and minimal electron neutrino contamination, with the capability of operating in anti-neutrino mode. The far detector with a 40 kt fiducial volume will use Liquid Argon Time Projection Chamber (LArTPC) technology for image-like precision in reconstructing neutrino interactions. We have used the latest DUNE files \cite{DUNE:2021cuw}, taking into account an exposure of 6.5 years in neutrino mode and 6.5 years in antineutrino mode to account for a total exposure of 312 kt-MW-years for each mode. The details of oscillation channels and systematic uncertainties are included in table \ref{tab:uncertainity}.

\subsection{T2HK} The Tokai to Hyper-Kamiokande (T2HK) \cite{Hyper-KamiokandeProto-:2015xww} experiment is a proposed long-baseline neutrino oscillation experiment designed to probe neutrino properties and CP violation. An intense neutrino beam produced at J-PARC will be directed towards the Hyper-Kamiokande (HyperK) detector, located 295 km away and 2.5$^\circ$ off-axis, with a planned beam power of 1.3 MW, corresponding to 2.7 $\times 10^{21}$ POT per year. The HyperK detector, an upgrade over Super-Kamiokande, has a cylindrical water Cherenkov module with a fiducial mass of 187 kilotons. Simulation studies assume this baseline and fiducial volume, with a 10-year data-taking period divided into 2.5 years in neutrino mode and 7.5 years in antineutrino mode. The channels and systematic uncertainties are as provided in table \ref{tab:uncertainity}.

\begin{table}[!h]
    \centering
    \begin{tabular}{|c|c|c|c|}
        \hline
        \multirow{2}{*}{Experiment details} &\multirow{2}{*}{Channels} & \multicolumn{2}{|c|}{Normalization error} \\\cline{3-4}
        & & Signal & Background \\
        \hline 
         \textbf{DUNE}, Baseline = 1300 km  &  & &\\
         L/E = 1543 km/GeV &  $\nu_e (\bar \nu_e)$ appearance & 2 \% (2\%) & 5 \% (5 \%) \\
          Fiducial mass = 40 kt (LArTPC) & $\nu_\mu (\bar \nu_\mu)$ disappearance & 5 \% (5 \%) & 5 \% (5 \%) \\ 
         Runtime = 6.5 yr $\nu + 6.5$ yr $\bar \nu$  &  & &\\
         \hline
         \textbf{T2HK}, Baseline = 295 km  &  & & \\
         L/E = 527 km/GeV &  $\nu_e (\bar \nu_e)$ appearance  & 3.6 \% (3.6 \%) & 10 \% (10 \%) \\ 
          Fiducial mass = 187 kt (WC)& $\nu_\mu (\bar \nu_\mu)$ disappearance  & 3.2 \% (3.9 \%) & 10 \% (10 \%)  \\ 
         Runtime = 2.5 yr $\nu + 7.5$ yr $\bar \nu$  &  & &\\
         \hline
    \end{tabular}
    \caption{Detector details and uncertainties for DUNE and T2HK.}
    \label{tab:uncertainity}
\end{table}

\subsection{Simulation details}

To evaluate the CP violation sensitivity of the LBL experiments, we calculate the $\Delta \chi^2$ using the pull method \cite{Fogli:2002pt}. We used the true values of the oscillation parameters listed in table \ref{tab:parameters}, which align with the current global best fit values \cite{Esteban:2024eli}. To account for systematic and statistical uncertainties, we marginalize over the atmospheric parameters $\theta_{23}$ and $\Delta m_{31}^2$ within their 3$\sigma$ ranges. Furthermore, to study the impact of dark NSI, moduli $|d_{\alpha \beta}|$ of the dark NSI elements and their associated phases are marginalized over the ranges $[0, 0.05]$ and $[-180^\circ, 180^\circ]$, respectively. All numerical simulations and statistical calculations are performed using the General Long Baseline Experiment Simulator (GLoBES)~\cite{Huber:2004ka, Huber:2007ji}, with the specific detector configurations and runtime detailed in table \ref{tab:uncertainity}.

The sensitivity of an experiment towards CP-violation is how well that particular experiment can discriminate between CP-conserving $(0, 180^\circ)$ and CP-violating values $(\neq 0, 180^\circ)$ of the $\delta_{CP}$ phase. The presence of dark NSI introduces additional sources of CP violation in the form of associated moduli $d_{\alpha \beta}$ and phases $\phi_{\alpha \beta}$, which can give rise to genuine or fake CP violation. Therefore, the $\chi^2$ for CP violation is defined as \cite{Masud:2016bvp, Medhi:2021wxj}, 

\be
\chi^2 \equiv \min_{\delta,|\eta|,\phi} \sum_{i=1}^{x} \sum_{j}^{2} \frac{\left[ N_{true}^{i,j}(\delta,|\eta|,\phi) - N_{test}^{i,j}(\delta=0,\pi; |\eta| \text{ range}; \phi) \right]^2}{N_{true}^{i,j}(\delta,|\eta|,\phi)},
\ee

\noindent where $N_{i,j}^{true}, N_{i,j}^{test}$ are the events in the true and test bins, respectively. To evaluate the sensitivity of a given experiment to CP-violating effects, the test value of the CP phase $\delta_{CP}$ is fixed at $0$ and $\pi$. The corresponding $\chi^2$ is then computed across the full range of true $\delta_{CP}$, i.e., $[-180^\circ, 180^\circ]$. In the resulting plots, the x-axis represents the variation in the true value of $\delta_{CP}$, while the vertical width of the gray bands reflects the variation in $\Delta \chi^2$ due to different true values of $\phi_{\alpha \beta}$ within their allowed ranges. Consequently, these shaded regions indicate the extrema of the $\Delta \chi^2$ for each true value of the CP-violating phase $\delta_{CP}$.

\begin{table}[!h]
    \centering
    \begin{tabular}{c|c|c|c|c|c|c}
        \hline
         Parameter  & $\theta_{23}$[$^\circ$]  & $\theta_{13}$[$^\circ$]   & $\theta_{12}$[$^\circ$]  & $\Delta m_{31}^2$[$ \times 10^{-3}$ eV]  & $\Delta m_{21}^2$ [$ \times 10^{-5}$ eV] & $\delta_{CP}$ [$^\circ$] \\
         \hline
         \hline
         Value & 47 & 8.62 & 33.45 & 2.53 & 7.49 & -90\\
         \hline 
         Range & [40-50] & Fixed & Fixed & [2.46-2.60]& Fixed & [-180,180]\\
        \hline
    \end{tabular}
    \caption{Values of the standard neutrino oscillation parameters used in our work. These values are consistent with current best-fit values \cite{Esteban:2020cvm, Esteban:2024eli}.}
    \label{tab:parameters}
\end{table}

\section{Result and discussions}\label{sec:result}
In this section, we first discuss the impact of dark NSI elements on oscillation probabilities, followed by an analysis of the CP-violation sensitivities at the long-baseline experiments DUNE and T2HK.

\subsection{Effect on the oscillation probabilities}

\begin{figure}[h!]
    \centering
     \includegraphics[width=.325\textwidth]{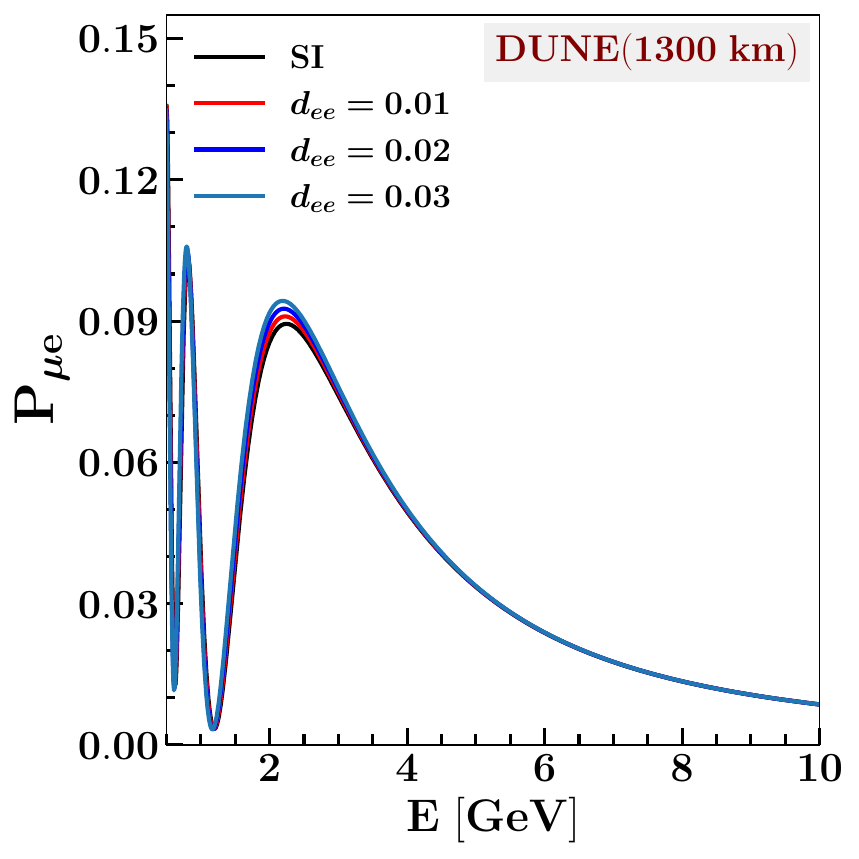}
    \includegraphics[width=.325\textwidth]{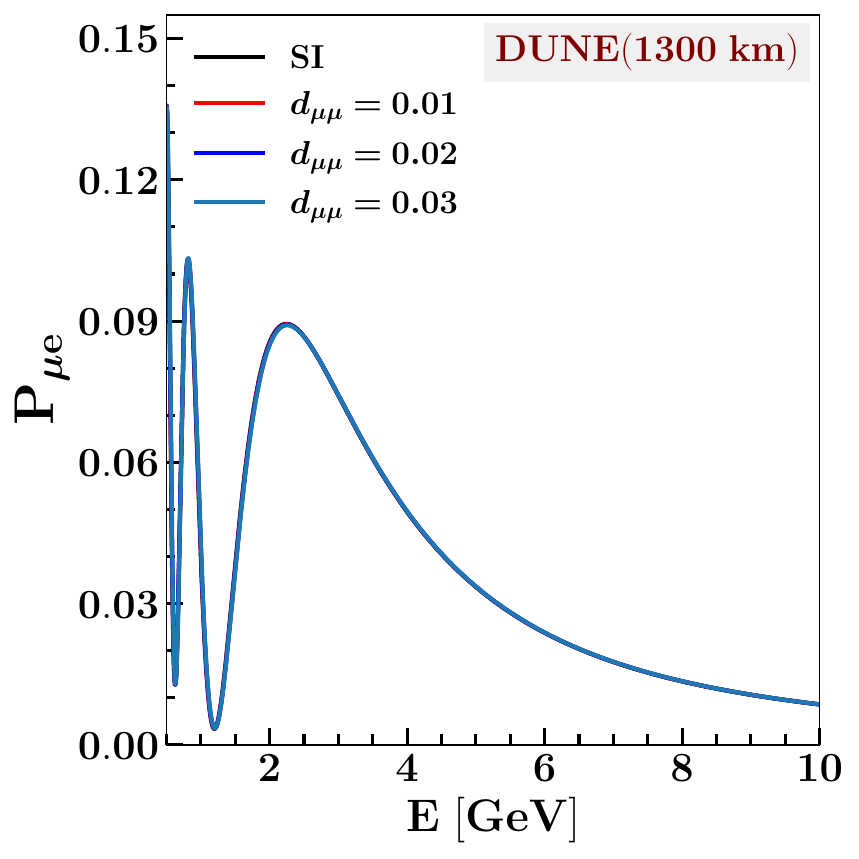}
    \includegraphics[width=.325\textwidth]{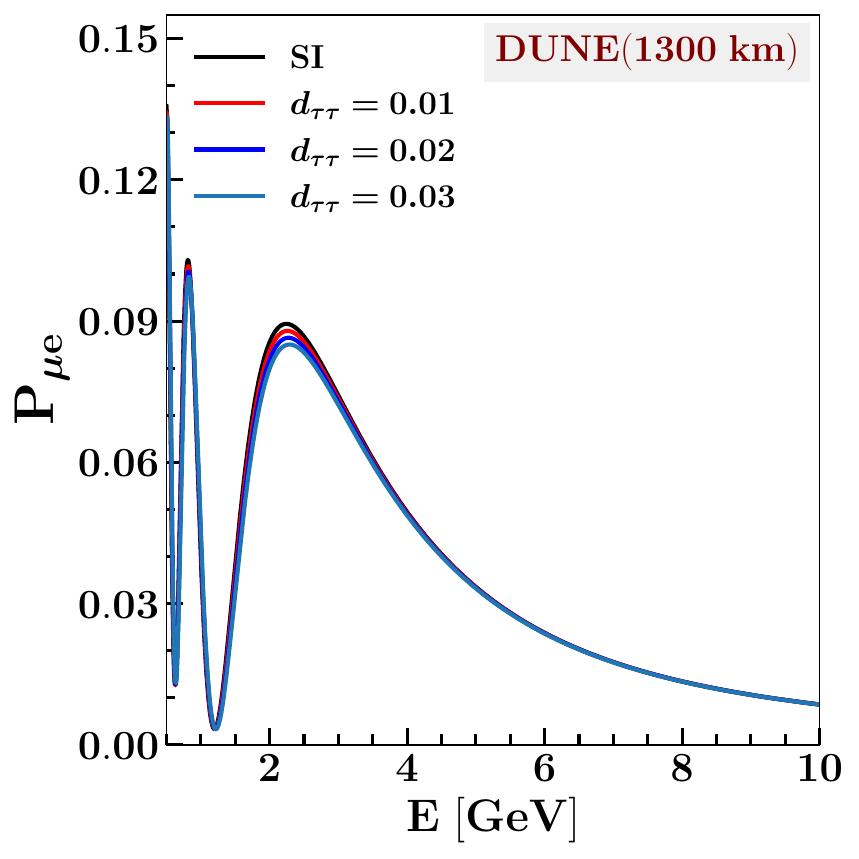}
     \includegraphics[width=.325\textwidth]{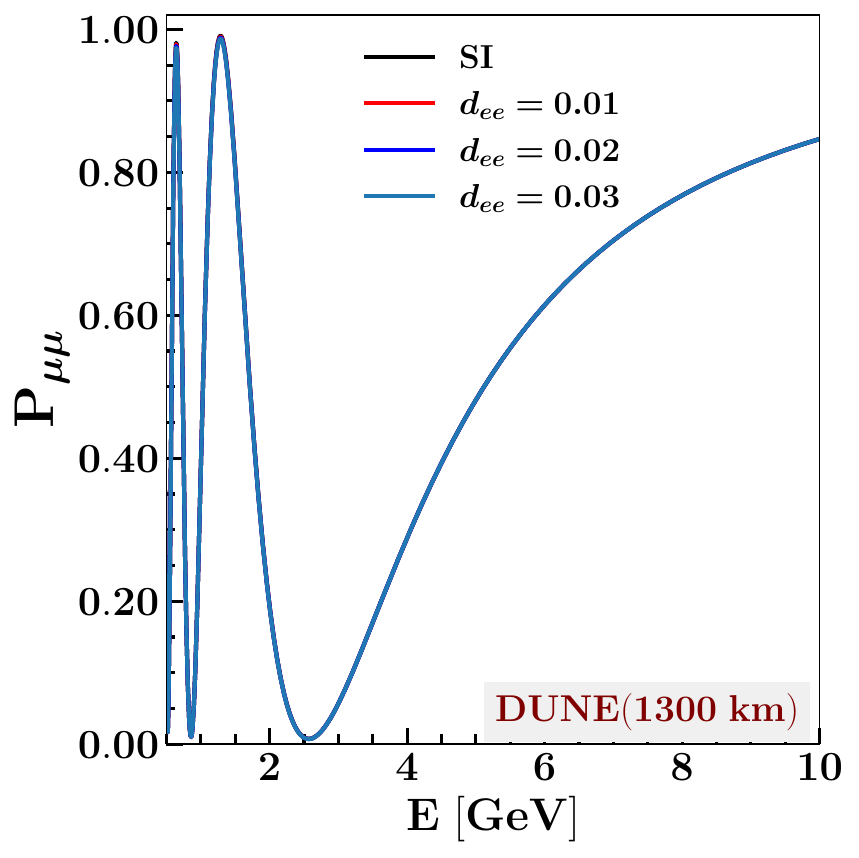}
    \includegraphics[width=.325\textwidth]{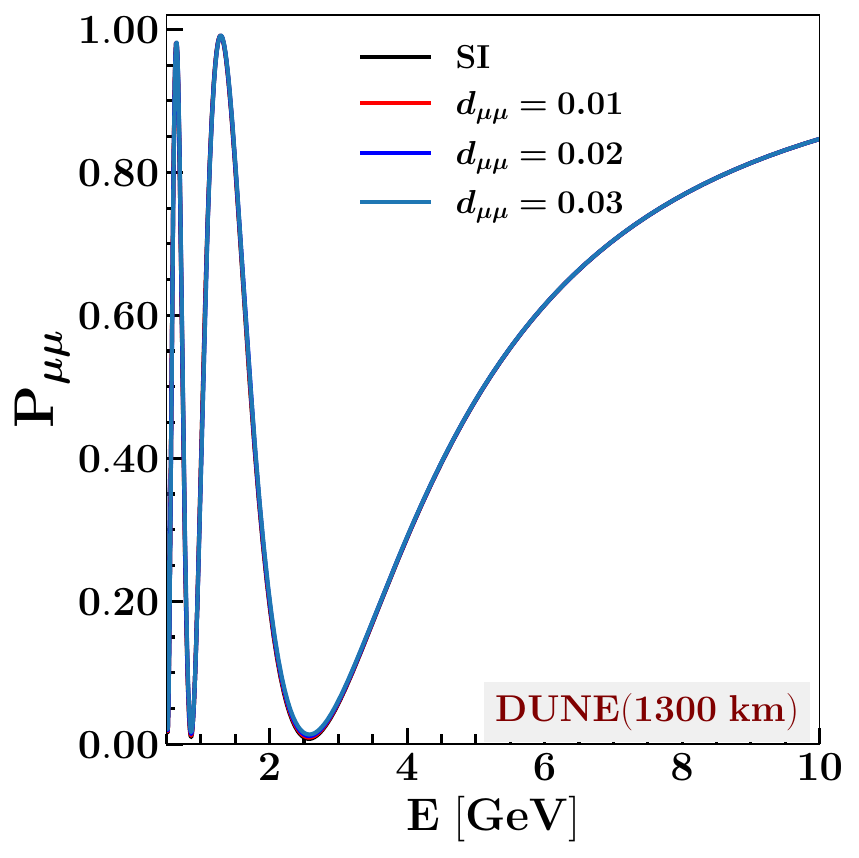}
    \includegraphics[width=.325\textwidth]{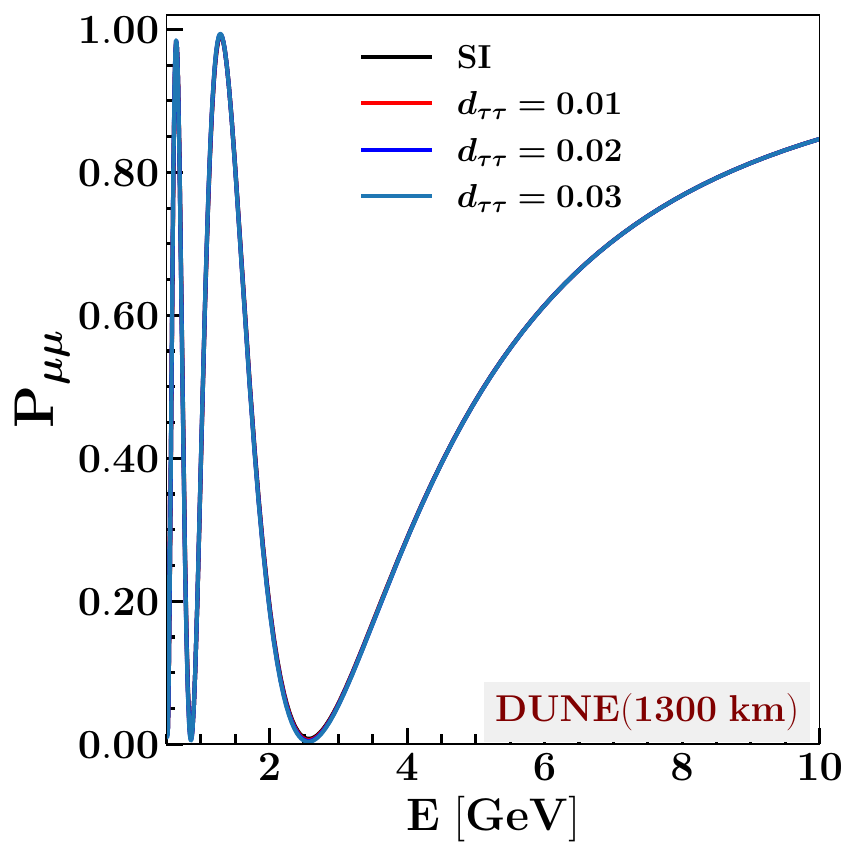}
    \caption{ Appearance probability $P_{\mu e}$ (top) and disappearance probability $P_{\mu \mu}$ (bottom) for DUNE ( L=1300 km), NO, $\delta_{CP}= -90^\circ$ for different values of the diagonal dark NSI parameters $\eta_{ee}$ (left), $\eta_{\mu \mu}$ (middle), and $\eta_{\tau \tau}$ (right).}
    \label{fig:prob_NO_DUNE_diag}
\end{figure}

\begin{figure}[h!]
    \centering
     \includegraphics[width=.325\textwidth]{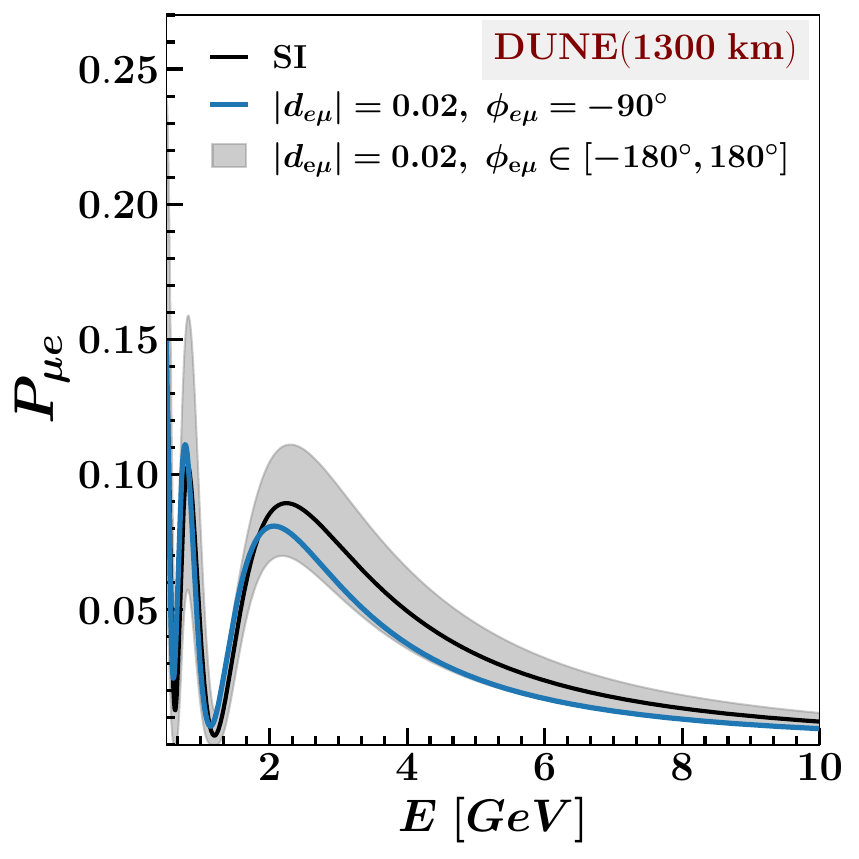}
    \includegraphics[width=.325\textwidth]{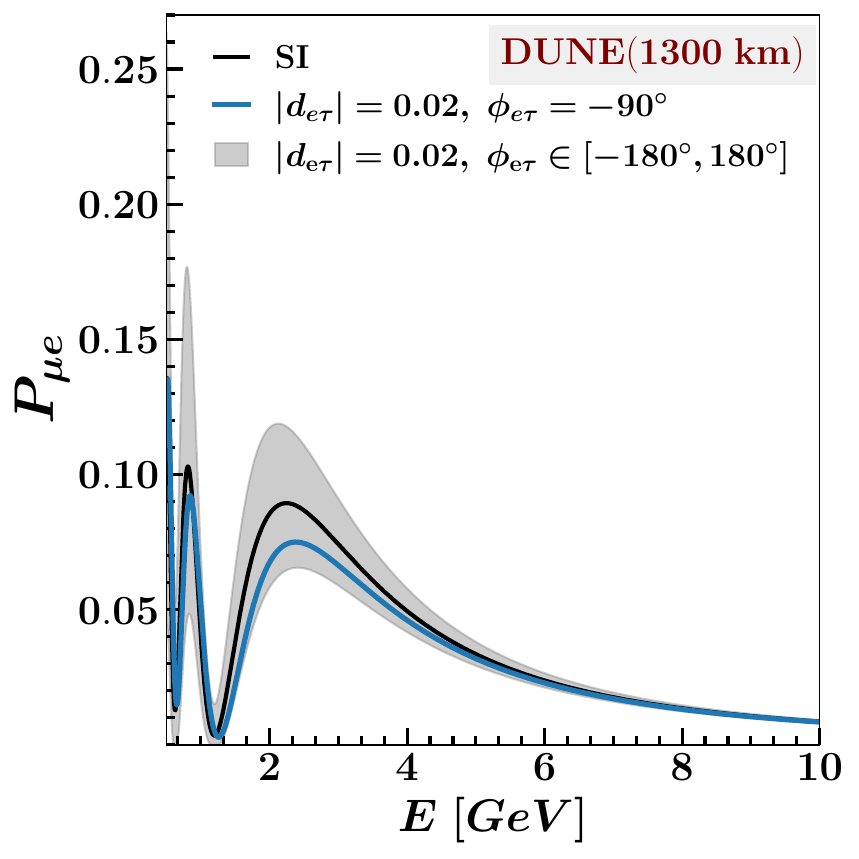}
    \includegraphics[width=.325\textwidth]{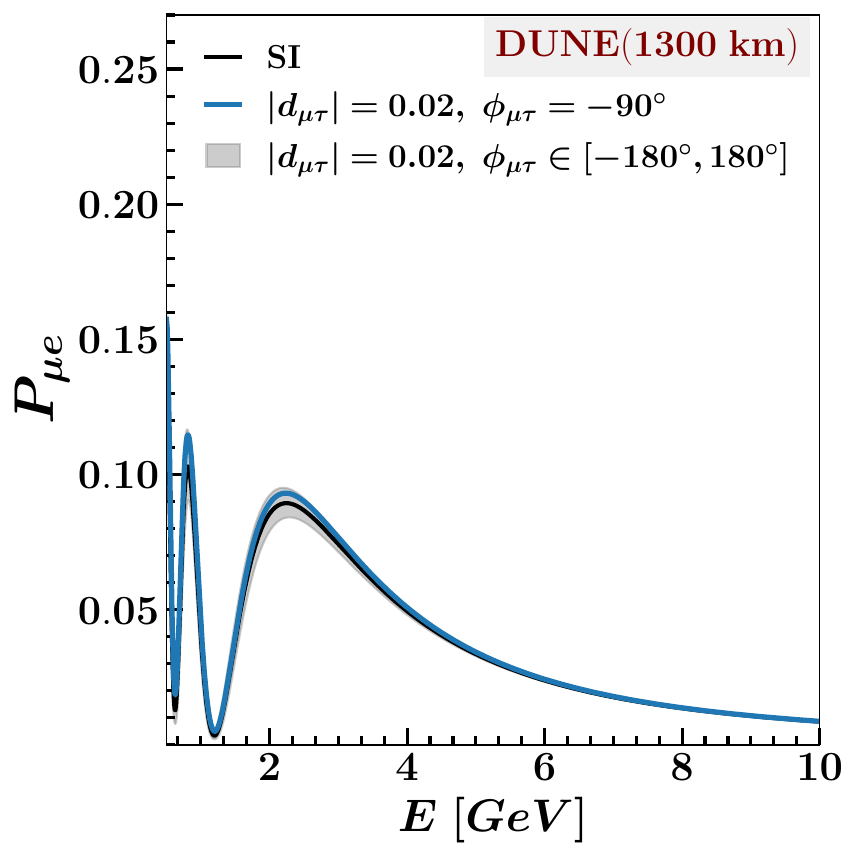}
    \includegraphics[width=.325\textwidth]{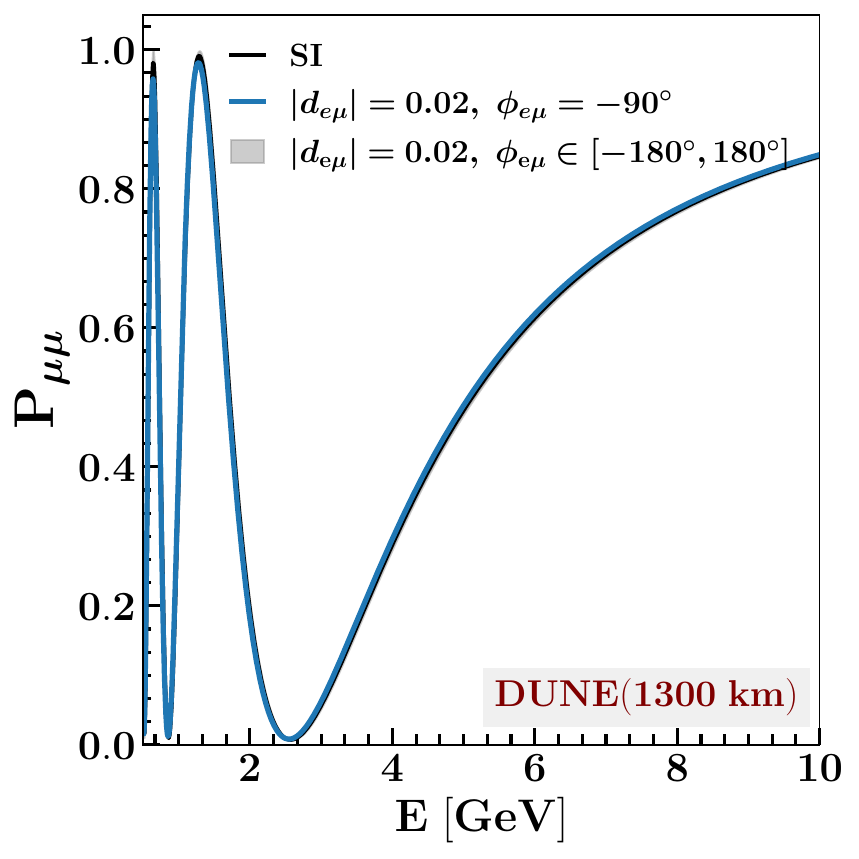}
    \includegraphics[width=.325\textwidth]{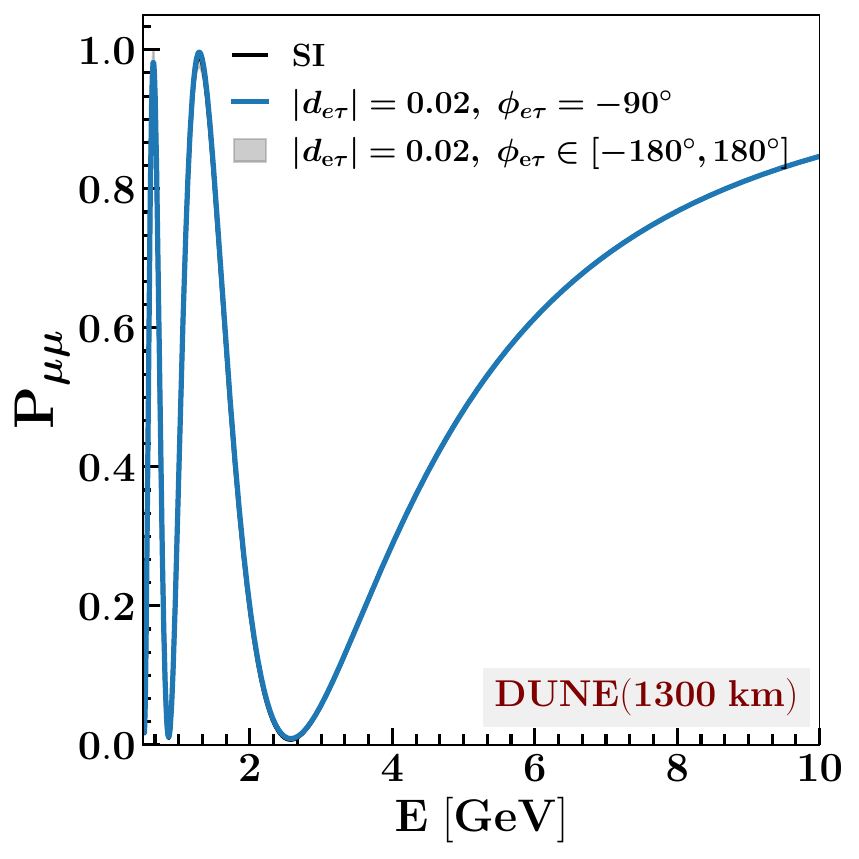}
    \includegraphics[width=.325\textwidth]{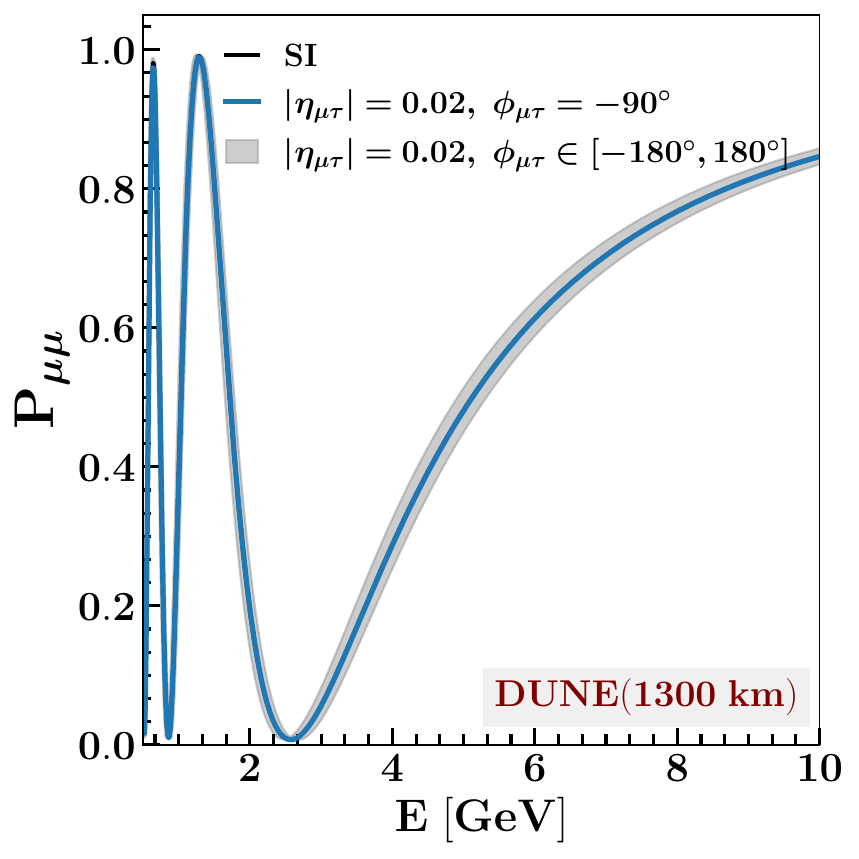}
    \caption{$P_{\mu e}$ (top), $P_{\mu \mu}$ (bottom) for DUNE ( L=1300 km), NO, $\delta_{CP}= -90^\circ,~\phi_{\alpha \beta}= 0^\circ$ for different values of the off-diagonal dark NSI parameters $\eta_{e\mu }$ (left), $\eta_{e \tau}$ (middle) $\eta_{\mu \tau}$ (right).}
    \label{fig:prob_NO_DUNE}
\end{figure}

In figures~\ref{fig:prob_NO_DUNE_diag} and \ref{fig:prob_NO_DUNE}, we show the impact of dark NSI elements on the appearance probability ($P_{\mu e}$) and disappearance probability ($P_{\mu \mu}$), respectively, for the 1300 km DUNE baseline. The oscillation parameters used in our analysis are listed in table~\ref{tab:parameters}. Figure~\ref{fig:prob_NO_DUNE_diag} illustrates the effect of the diagonal NSI elements. The black line corresponds to the Standard Interaction (SI) case (i.e., no dark NSI, $d_{\alpha\beta} = 0$). The red, dark blue, and light blue lines correspond to the values of the diagonal NSI elements $d_{\alpha \alpha}$, as indicated in the legend. Figure~\ref{fig:prob_NO_DUNE} shows the effect of the off-diagonal dark NSI elements and their associated phases. The solid black and blue lines represent the SI case and the $d_{\alpha\beta} = 0.02$, $\phi_{\alpha\beta}= -90^\circ$ case, respectively. The gray band is generated by varying the off-diagonal phase $\phi_{\alpha\beta}$ within the range $[-180^\circ, 180^\circ]$ while fixing $d_{\alpha\beta} = 0.02$. The outcomes of this analysis are listed below,

\begin{itemize}

    \item The diagonal dark NSI elements $d_{\mu \mu}$ (top middle) has a nominal impact on the appearance channel. In the presence of $d_{ee}$ (top left), there is a slight enhancement of $P_{\mu e}$ around the oscillation maxima. Whereas in presence of $d_{\tau \tau}$ (top right), there is marginal suppression of $P_{\mu e}$ as that of the SI case. All three diagonal dark NSI parameters $d_{ee}$, $d_{\mu \mu}$, $d_{\tau \tau}$ have a nominal impact on $P_{\mu \mu}$ (bottom panel).
    
    \item For the off-diagonal element $d_{e\mu}$ (top left), when the phase $\phi_{e \mu}$ is $-90^\circ$, we observe a suppression in the oscillation probability and also a slight shift in the position of the peak to the left. The dark NSI phase, $\phi_{e \mu}$ can significantly affect the oscillation probabilities. There may be a suppression or enhancement of $P_{\mu e}$ depending on the value of $\phi_{e \mu}$.

    \item In the presence of $d_{e \tau}$ (top middle), we observe a suppression in the oscillation probability around oscillation maxima when $\phi_{e \tau} = -90^\circ$. Similar to $\phi_{e \mu}$, the dark NSI phase $\phi_{e \tau}$ can also significantly affect the oscillation probability.

    \item We observe a nominal suppression in appearance probability in presence of $d_{\mu \tau}$ (top right) when $\phi_{\mu \tau}= -90^\circ$. The effect of the dark NSI phase $\phi_{\mu \tau}$ also has a marginal effect on $P_{\mu e}$. 
    
    \item The off-diagonal elements $d_{e\mu}$, $d_{e\tau}$, and $d_{\mu\tau}$ (bottom panel) have nominal impact on the $P_{\mu \mu}$ disappearance channel.

\end{itemize}

\begin{figure}[h!]
    \centering
     \includegraphics[width=.325\textwidth]{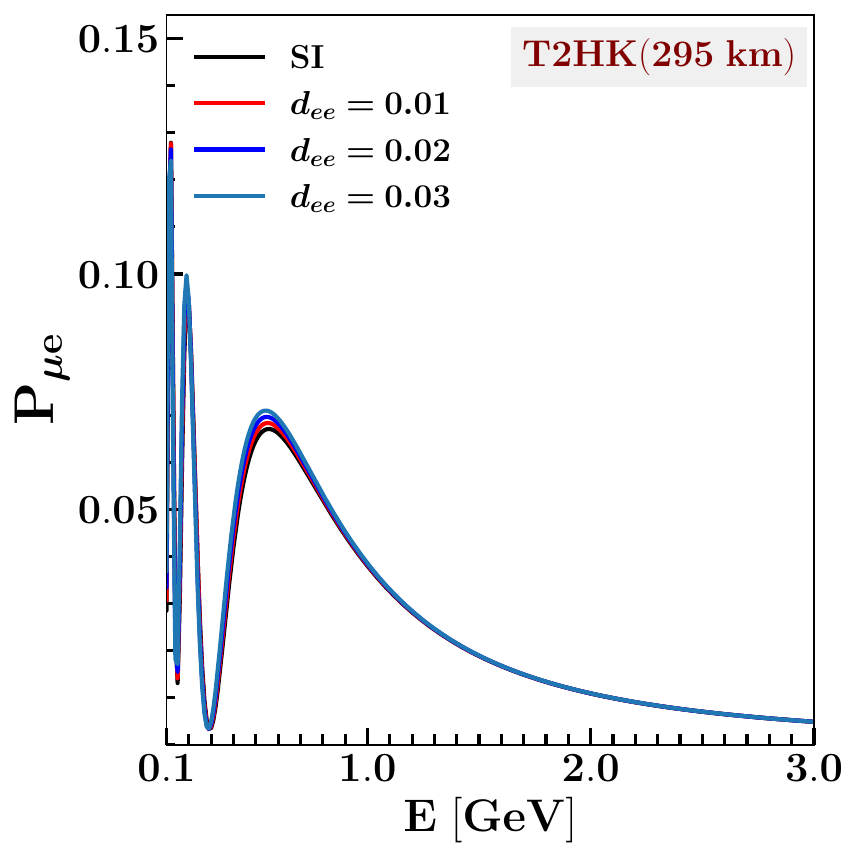}
    \includegraphics[width=.325\textwidth]{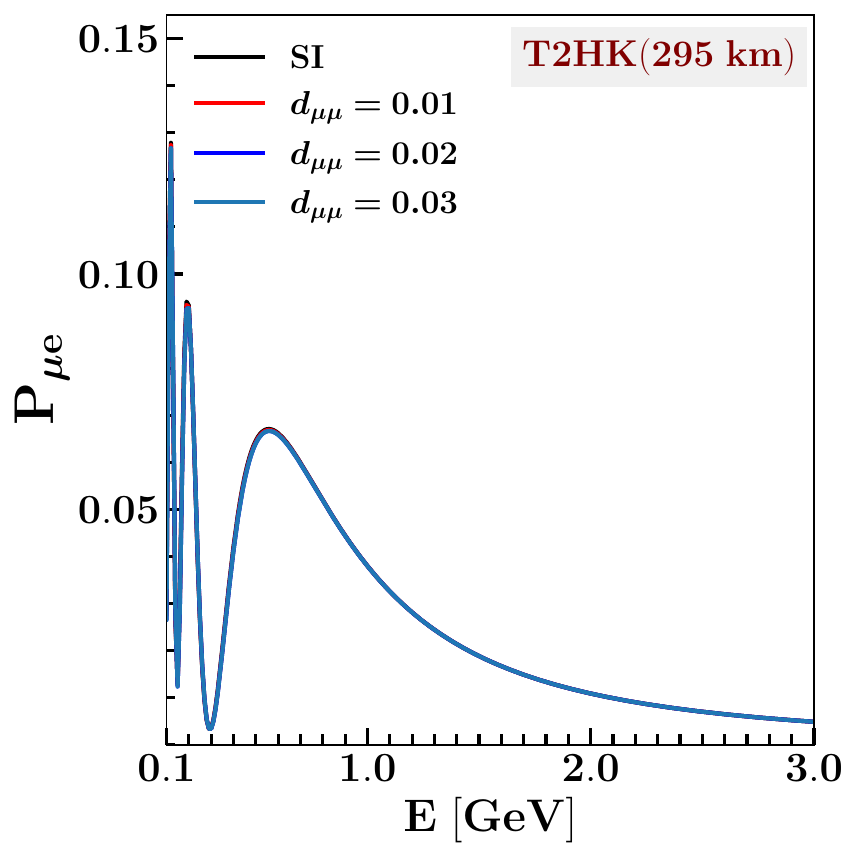}
    \includegraphics[width=.325\textwidth]{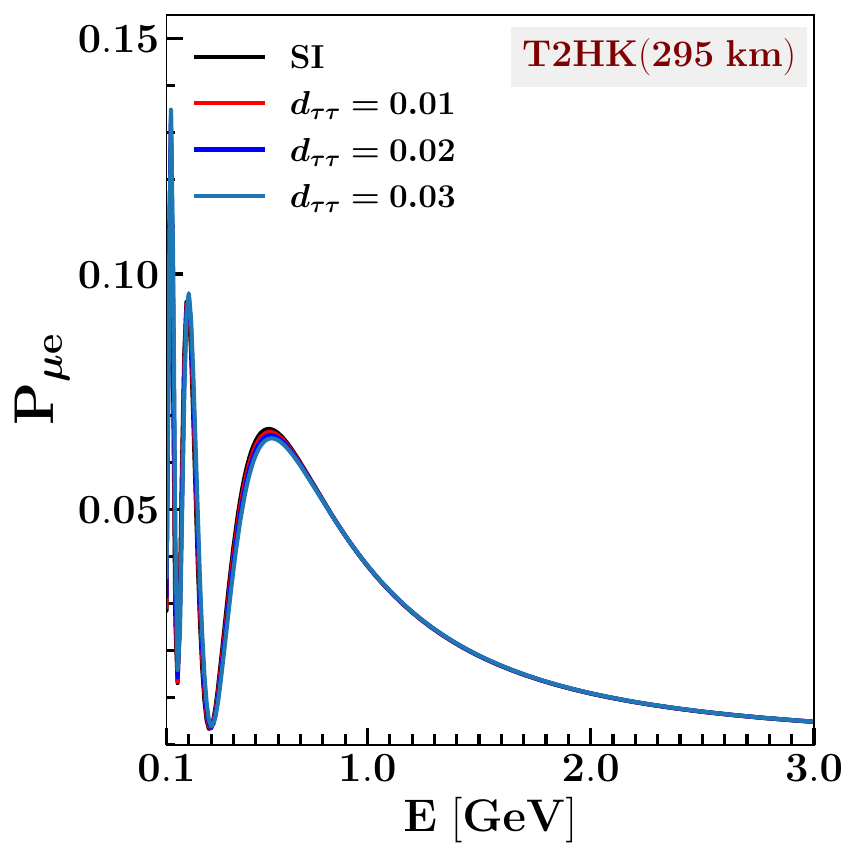}
     \includegraphics[width=.325\textwidth]{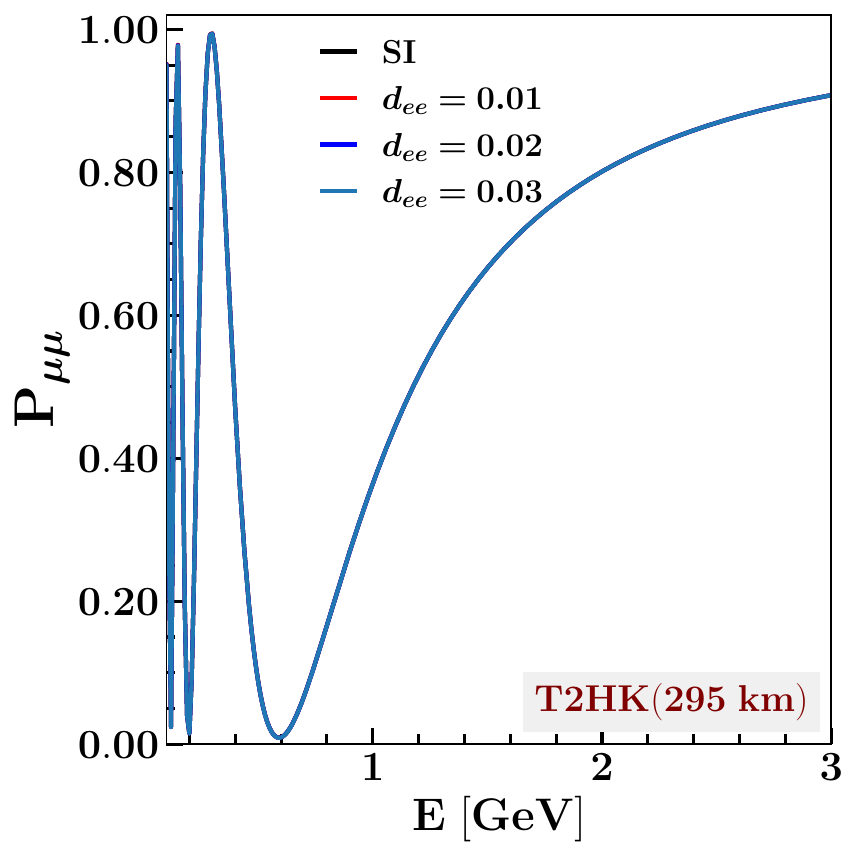}
    \includegraphics[width=.325\textwidth]{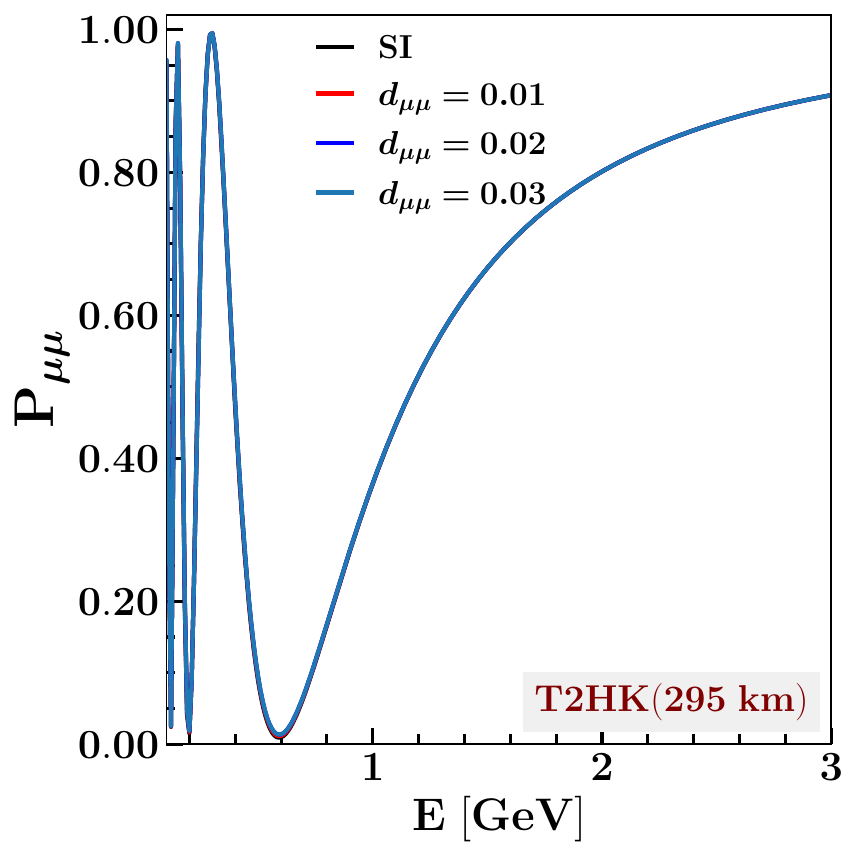}
    \includegraphics[width=.325\textwidth]{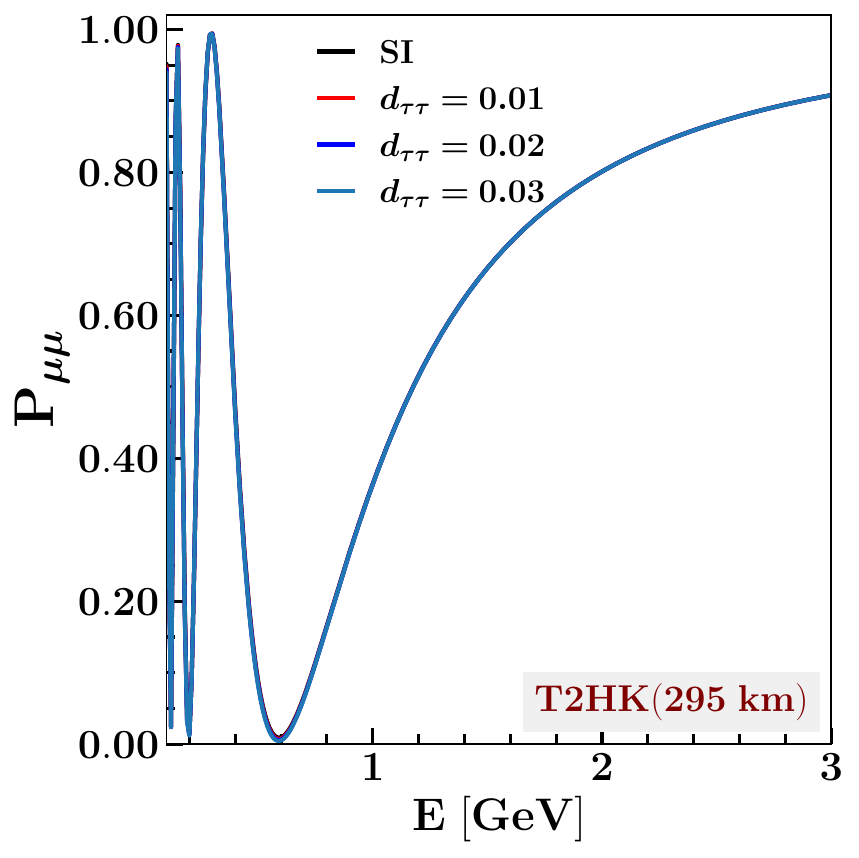}
    \caption{Appearance probability $P_{\mu e}$ (top) and disappearance probability $P_{\mu \mu}$ (bottom) for T2HK ( L=295 km), NO, $\delta_{CP}= -90^\circ$ for different values of the diagonal dark NSI parameters $\eta_{ee}$ (left), $\eta_{\mu \mu}$ (middle), and $\eta_{\tau \tau}$ (right).}
    \label{fig:prob_NO_T2HK_diag}
\end{figure}

\begin{figure}[h!]
    \centering
     \includegraphics[width=.325\textwidth]{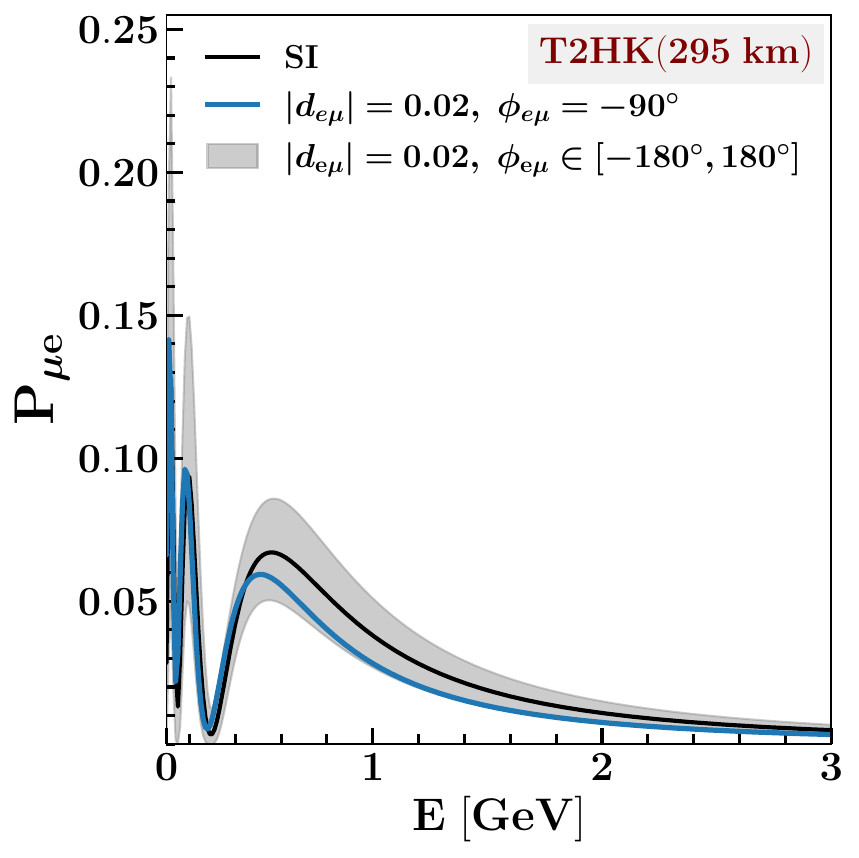}
    \includegraphics[width=.325\textwidth]{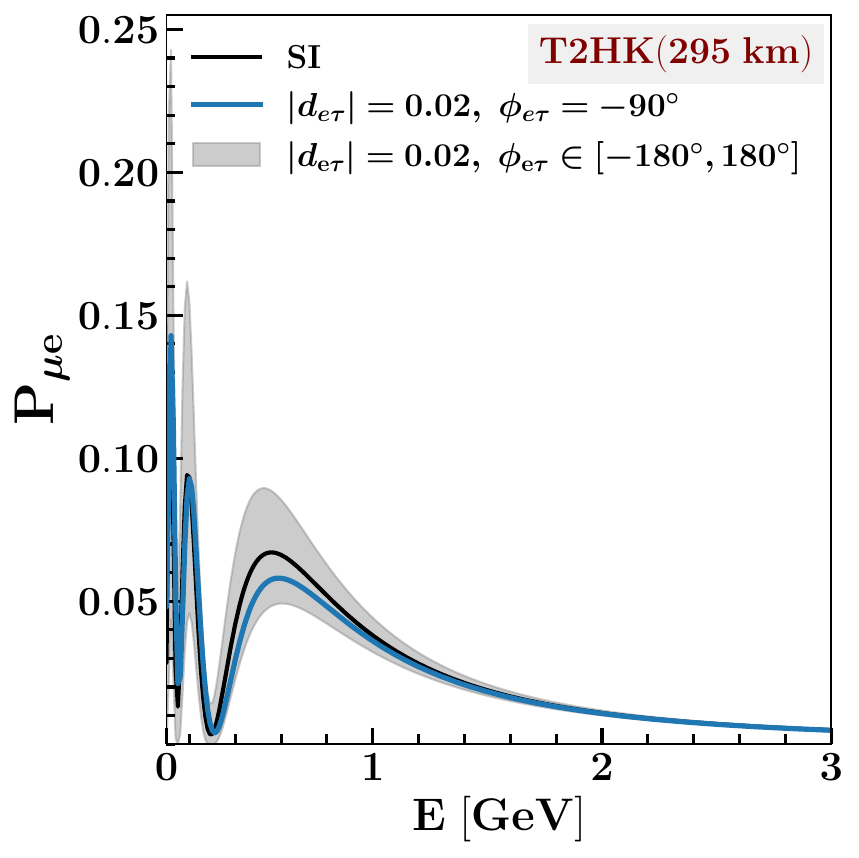}
    \includegraphics[width=.325\textwidth]{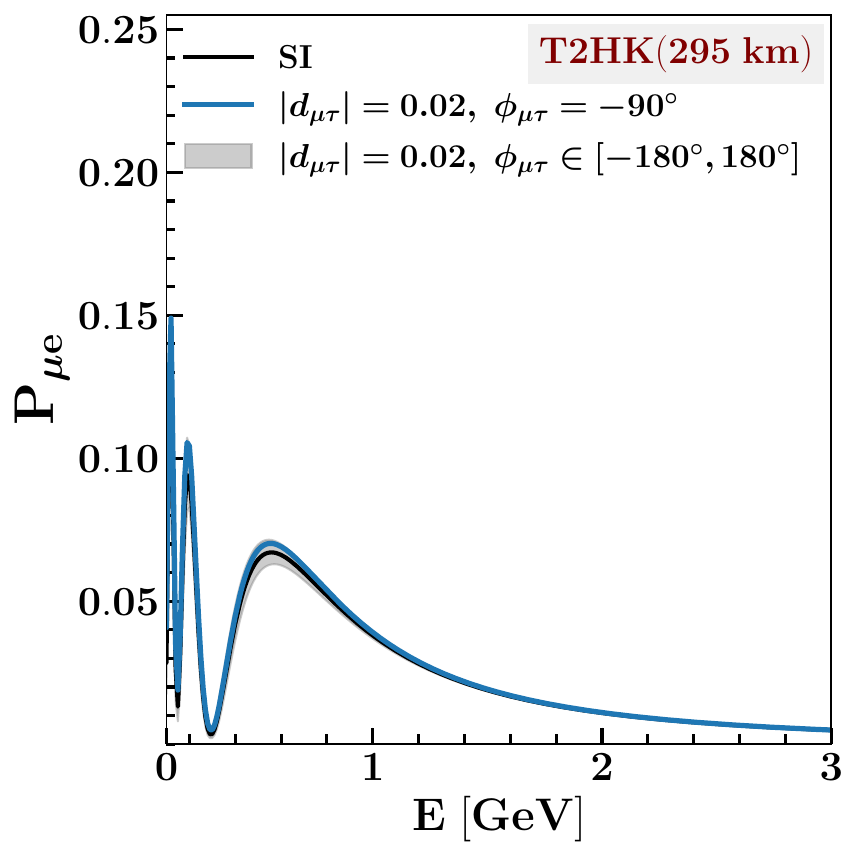}
      \includegraphics[width=.325\textwidth]{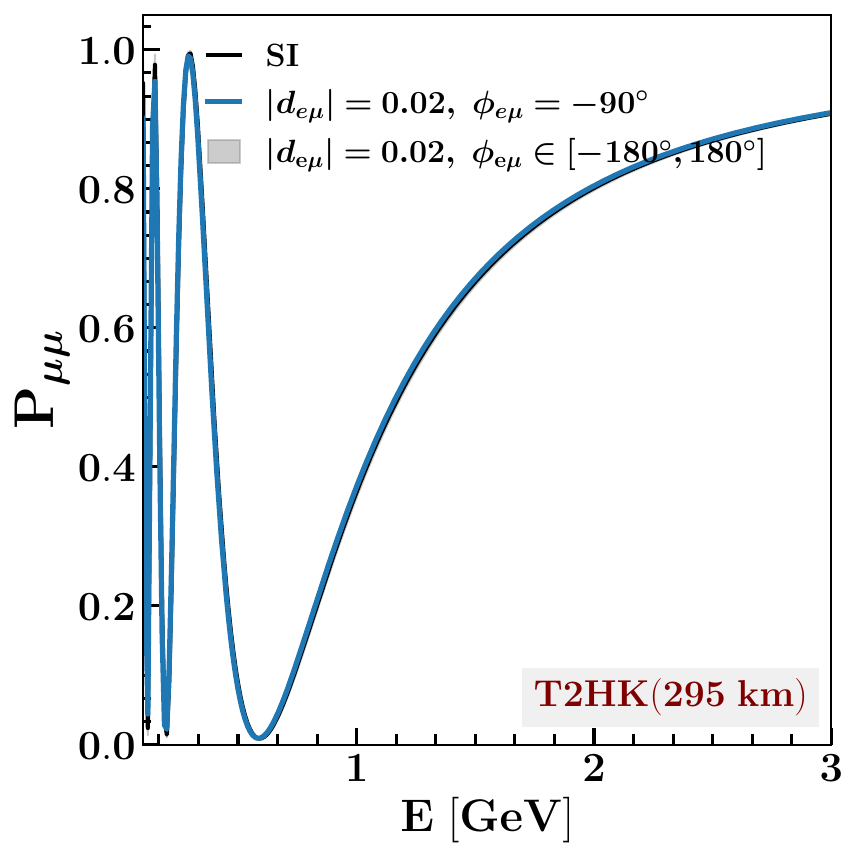}
    \includegraphics[width=.325\textwidth]{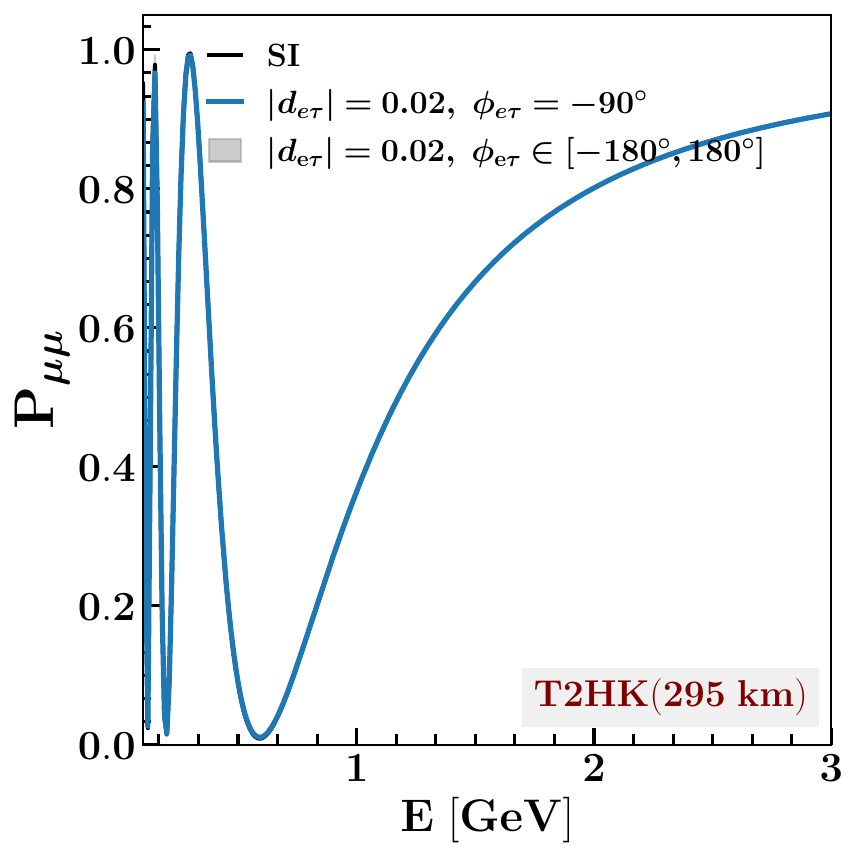}
    \includegraphics[width=.325\textwidth]{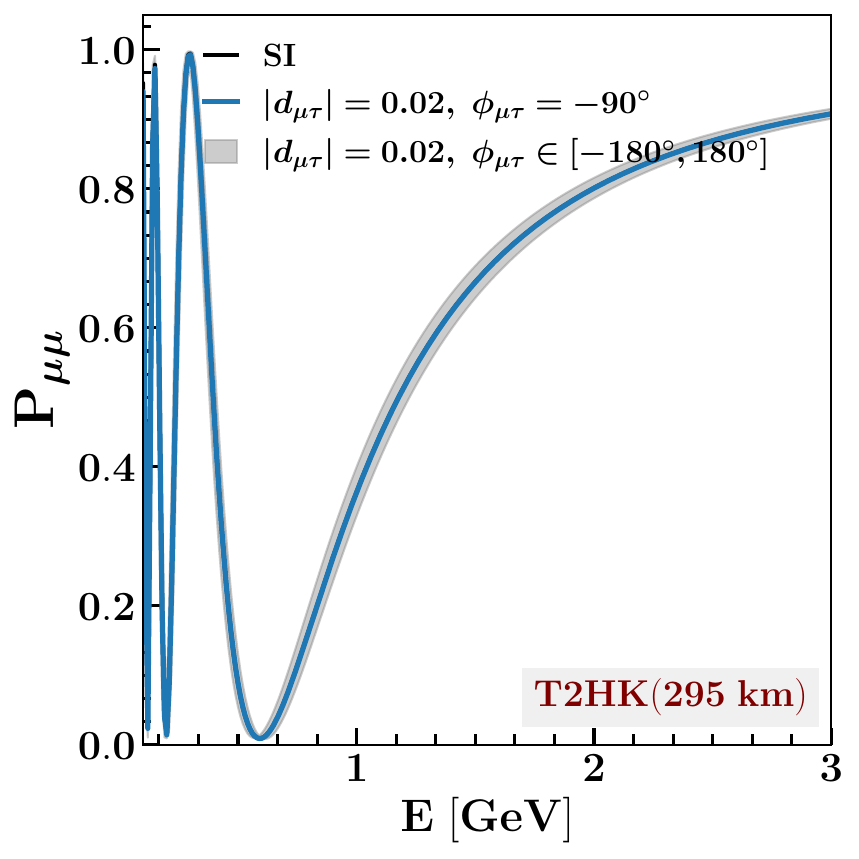}
    \caption{$P_{\mu e}$ (top), $P_{\mu \mu}$ (bottom) for T2HK ( L=295 km), NO, $\delta_{CP}= -90^\circ,~\phi_{\alpha \beta}= 0^\circ$ for different values of the off-diagonal dark NSI parameters $\eta_{e\mu }$ (left), $\eta_{e \tau}$ (middle) $\eta_{\mu \tau}$ (right).}
    \label{fig:prob_NO_T2HK}
\end{figure}

Figures~\ref{fig:prob_NO_T2HK_diag} and \ref{fig:prob_NO_T2HK} illustrate the impact of diagonal and off-diagonal dark NSI elements on $P_{\mu e}$ and $P_{\mu \mu}$, respectively for the 295 km T2HK baseline. For this shorter baseline, the effect of the diagonal dark NSI parameters ($d_{ee}$, $d_{\mu \mu}$, $d_{\tau \tau}$) is also found to be nominal. The detailed observations from these figures are as follows,

\begin{itemize}
    \item For the T2HK baseline, the diagonal dark NSI parameters $d_{ee}$, $d_{\mu\mu}$, and $d_{\tau\tau}$ are found to have a negligible influence on both the appearance probability $P_{\mu e}$ and the disappearance probability $P_{\mu\mu}$. This indicates that, within the chosen benchmark values, diagonal dark NSI effects do not significantly modify the standard oscillation patterns at the relatively short T2HK baseline, where matter effects are inherently weaker.

    \item The off-diagonal dark NSI parameters exhibit a more pronounced impact on the appearance channel. In particular, for the chosen values of $d_{e\mu}$ (top left) and $d_{e\tau}$ (top middle),  $P_{\mu e}$ is notably suppressed when the associated CP phases $\phi_{\alpha\beta}$ are set to $-90^\circ$, consistent with behavior observed at the DUNE baseline. In contrast, the parameter $d_{\mu\tau}$ (top right) induces only a marginal modification in $P_{\mu e}$. Furthermore, the phases $\phi_{e\mu}$ and $\phi_{e\tau}$ play a significant role in shaping the oscillation probability, while the effect of $\phi_{\mu\tau}$ remains comparatively subdominant.

    \item In the disappearance channel, $P_{\mu\mu}$ (bottom pane), the sensitivity to off-diagonal dark NSI parameters is considerably reduced. The parameters $d_{e\mu}$, $d_{e\tau}$, and $d_{\mu\tau}$, along with their associated CP phases $\phi_{\alpha\beta}$, introduce only minimal deviations from the standard oscillation scenario.
\end{itemize}

In order to quantitatively explore the impact of dark NSI on $P_{\mu e}$ over the $\delta_{CP}$ parameter space, we define a quantity $\Delta P_{\mu e}$ as

\begin{equation}
  \rm \Delta P_{\mu e} =|\rm P_{\mu e}^{DNSI}-P_{\mu e}^{SI}|.  
\end{equation}

\noindent Here, $\rm P_{\mu e}^{DNSI}$ is the 
appearance probability in the presence of dark NSI and $P_{\mu e}^{SI}$ is the appearance probability without dark NSI. 

\begin{figure}[h!]
    \centering
    \includegraphics[width=0.32\linewidth]{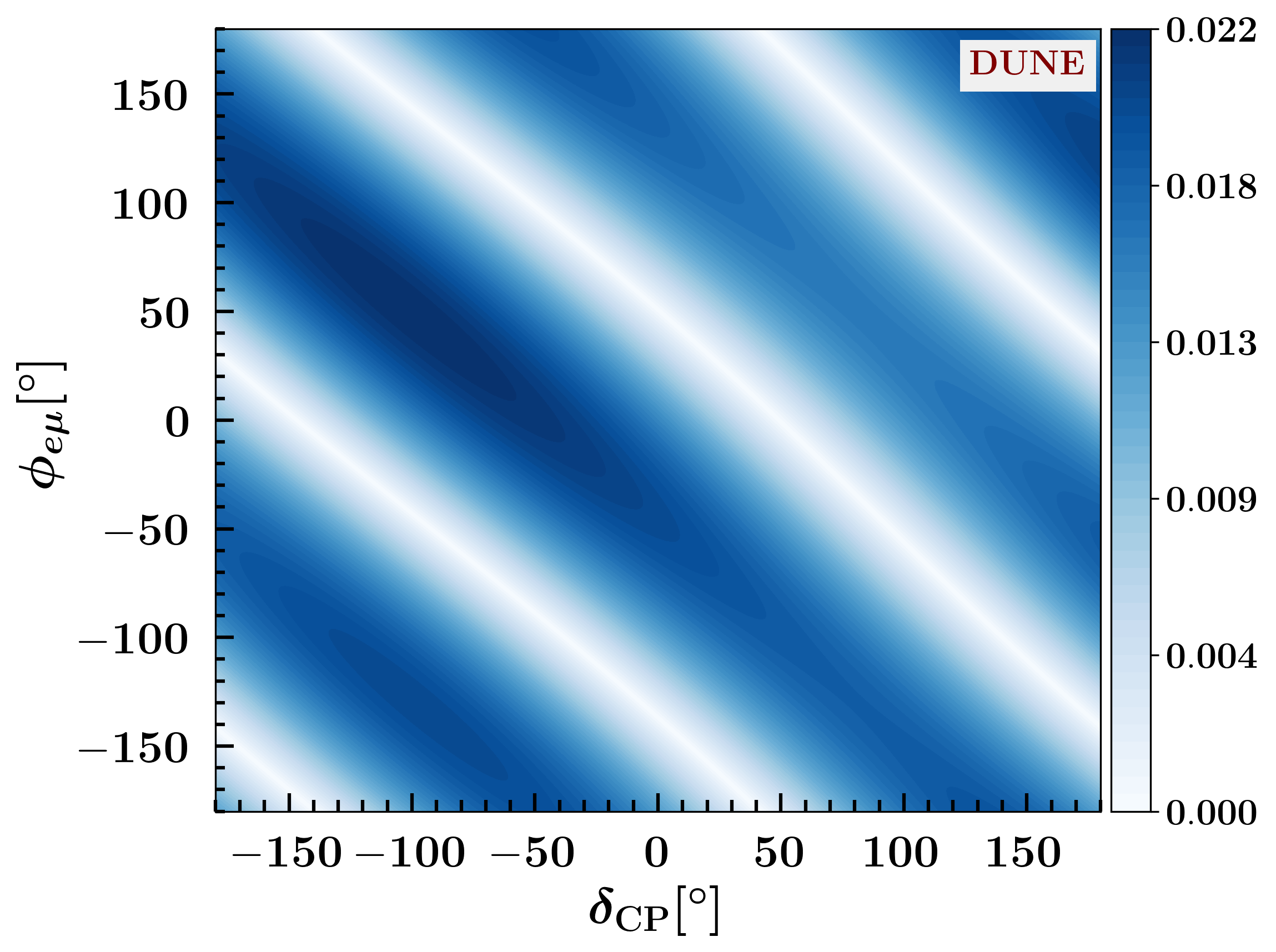}
    \includegraphics[width=0.32\linewidth]{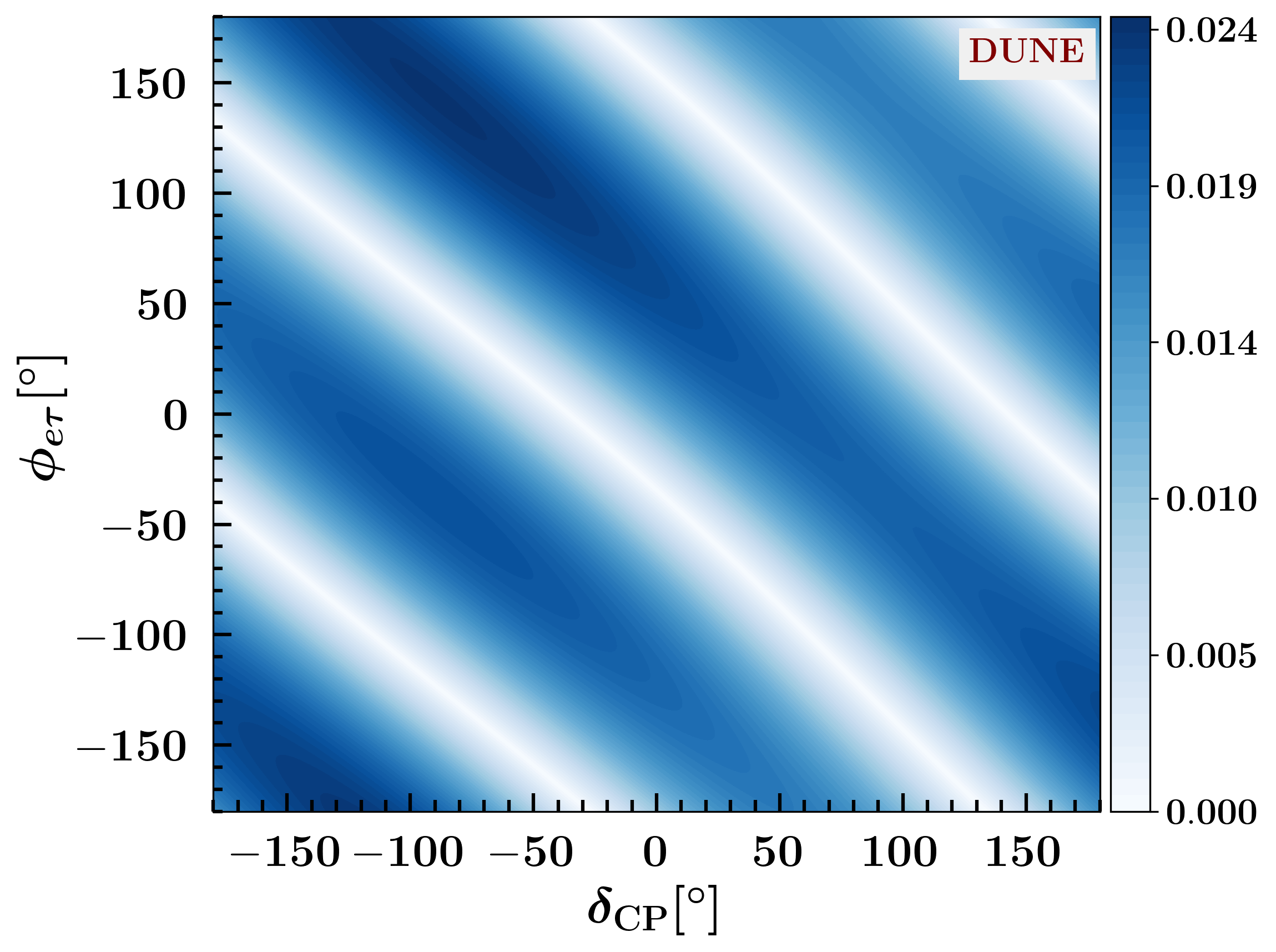}
    \includegraphics[width=0.32\linewidth]{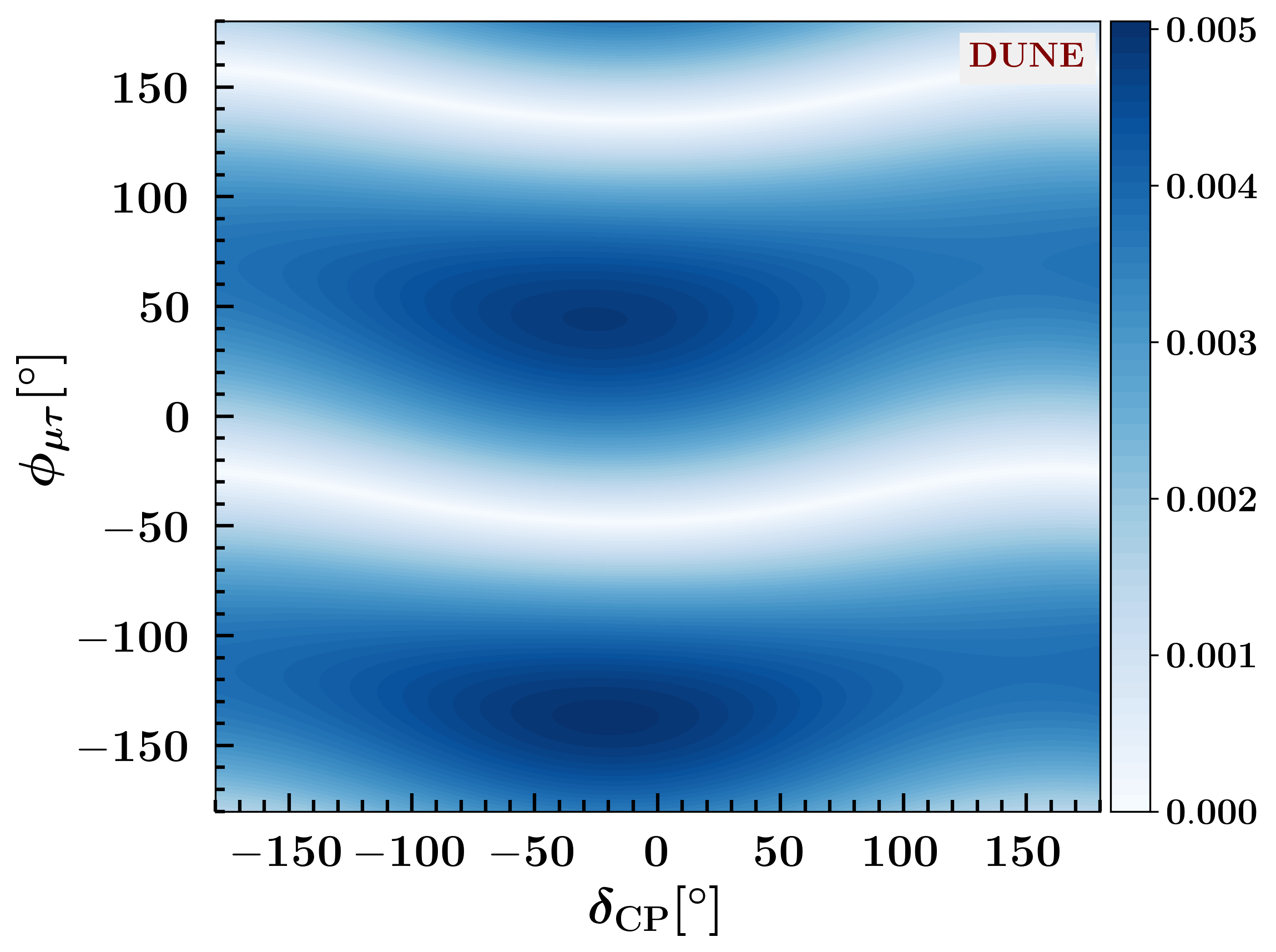}
    \includegraphics[width=0.32\linewidth]{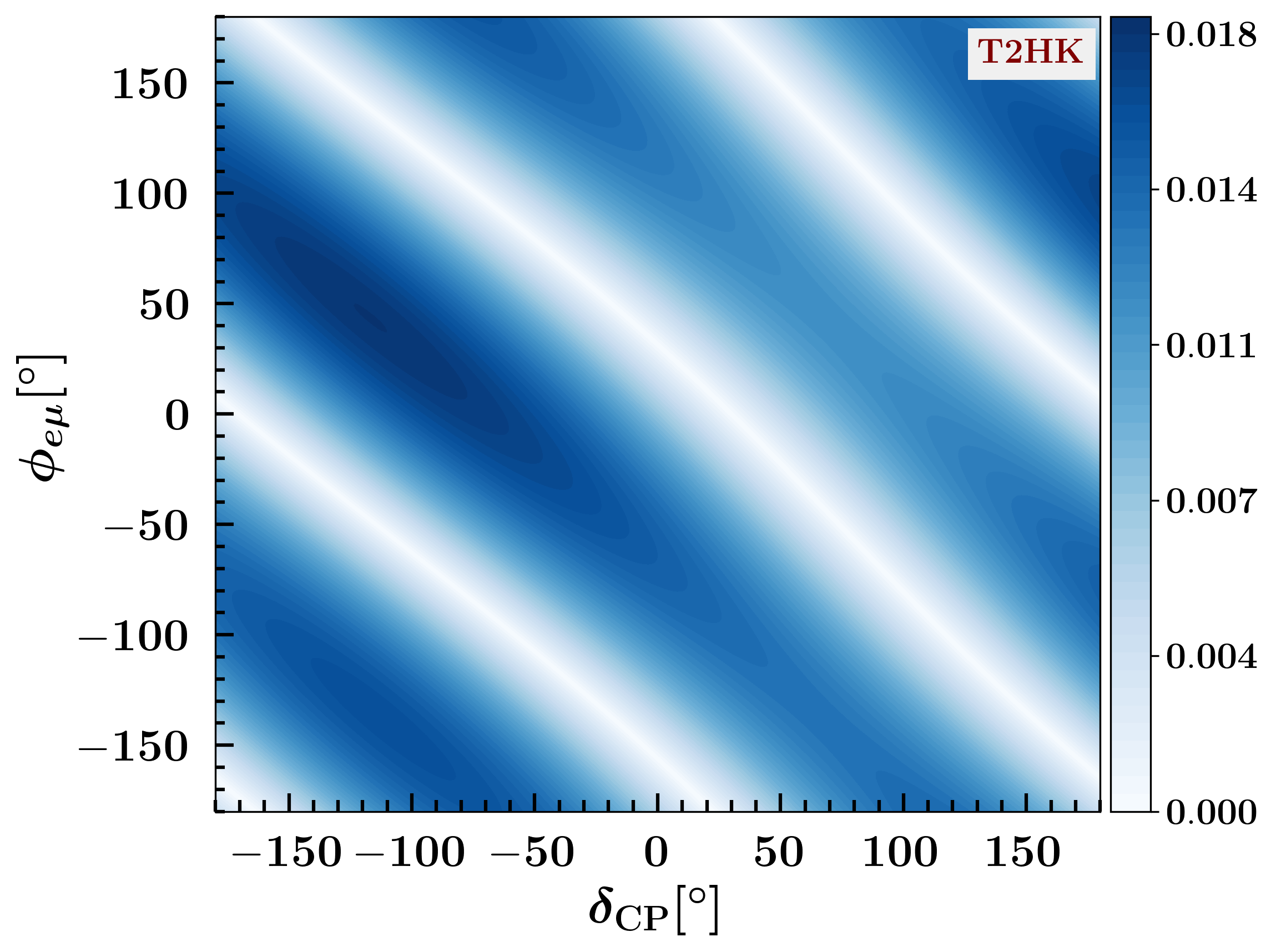}
    \includegraphics[width=0.32\linewidth]{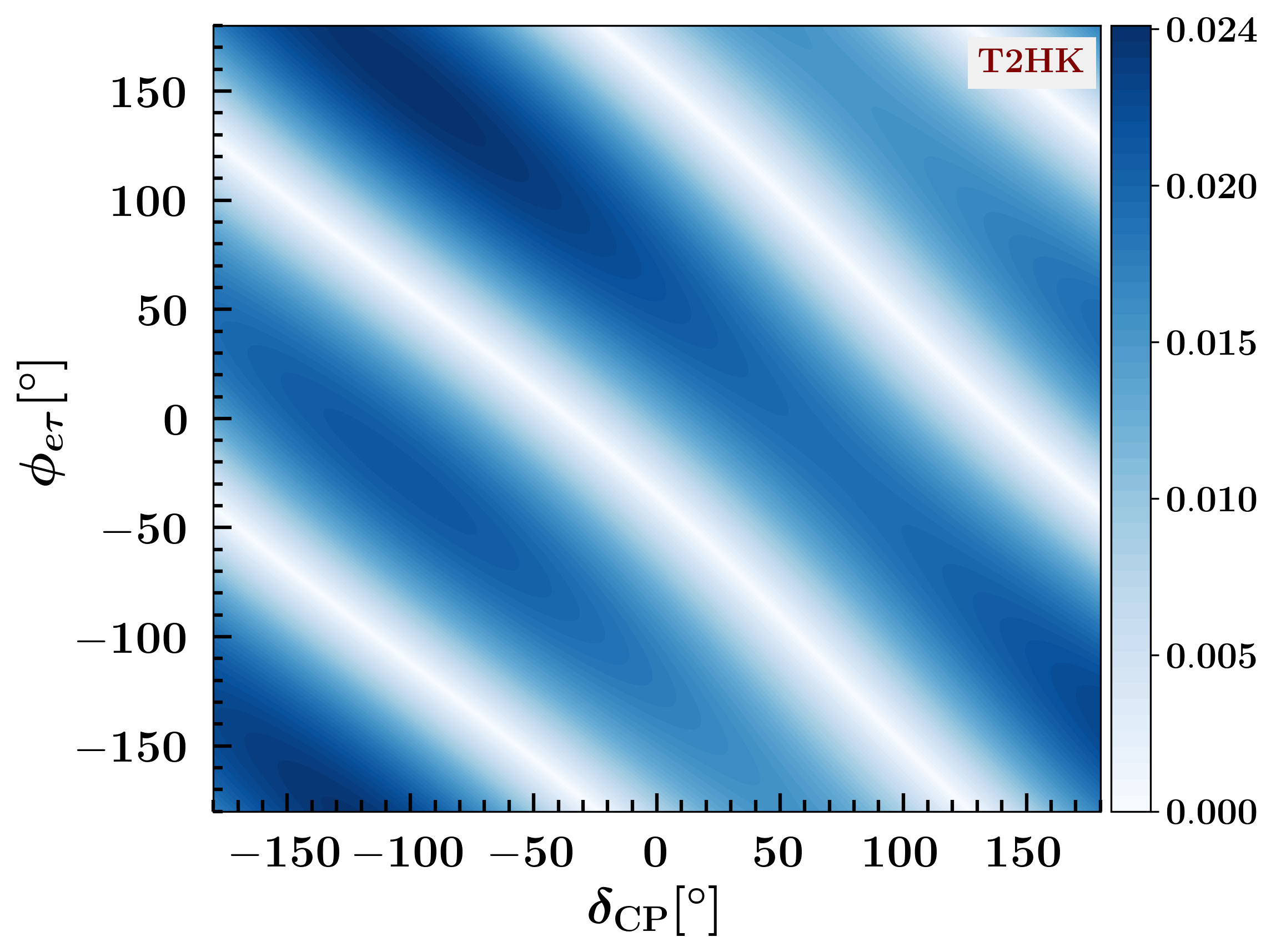}
    \includegraphics[width=0.32\linewidth]{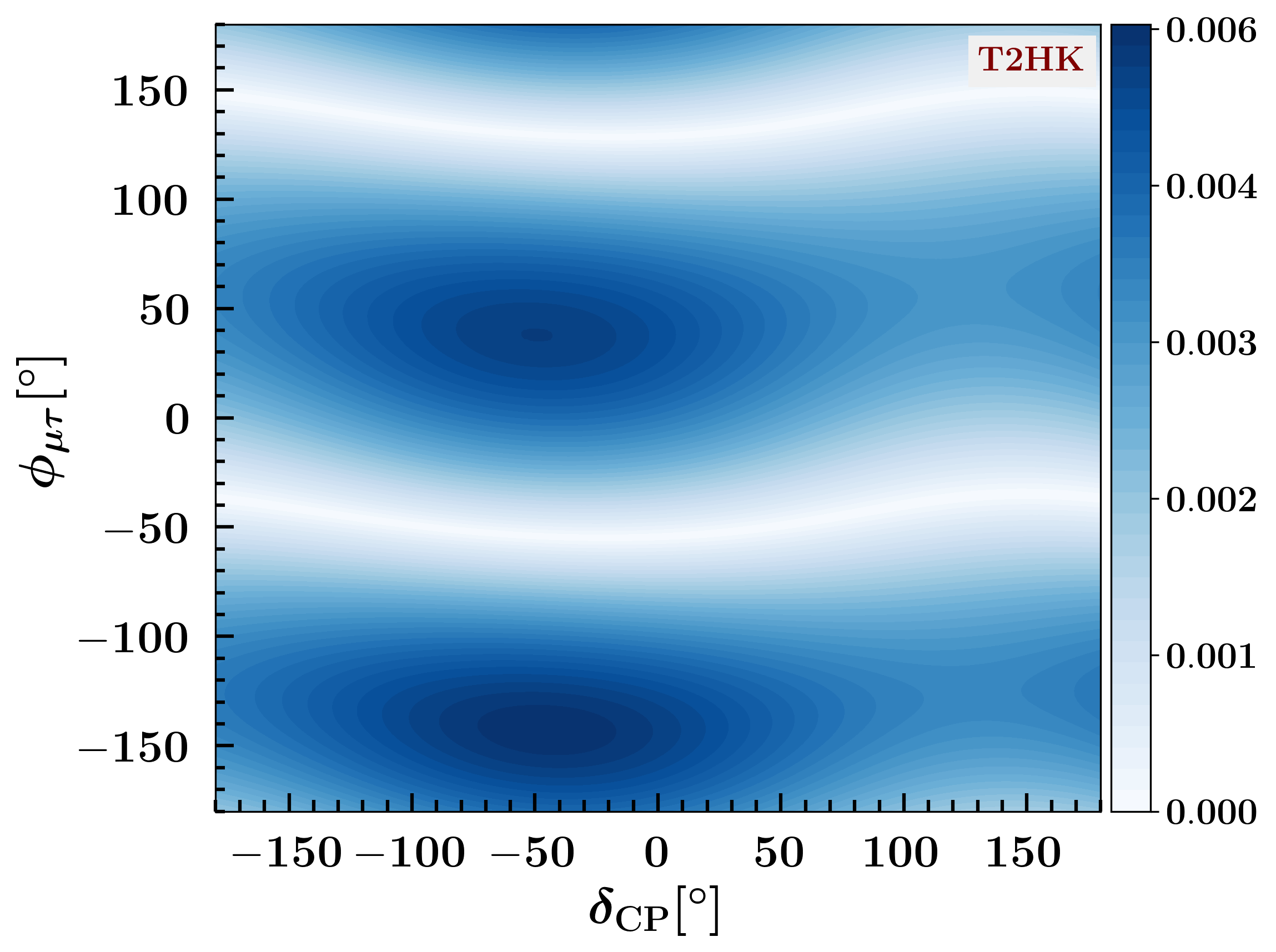}
    \caption{$\Delta P_{\mu e} =|\rm P_{\mu e}^{DNSI}-P_{\mu e}^{SI}|$ in $\phi_{\alpha \beta}$ vs $\delta_{CP}$ plane for DUNE (top--panel) and T2HK (bottom--panel) for $d_{\alpha \beta} = 0.02$. For DUNE, the peak energy is fixed at $\rm E \sim 2.5~GeV$, and for T2HK it is fixed at $\rm E \sim 0.5~GeV$. }
    \label{fig:delP}
\end{figure}

In figure~\ref{fig:delP}, we study the impact of ($\phi_{e\mu}$, $\phi_{e\tau}$, $\phi_{\mu\tau}$) with $\delta_{CP}$ for both DUNE (top-panel) and T2HK (bottom-panel). We examine the effects of dark NSI on $P_{\mu e}$ for the whole $\delta_{CP}$ parameter space to look for potential regions of degeneracies at the probability level. For DUNE, we have fixed the energy  $E\sim$ 2.5 GeV, and for T2HK, we fixed it at $E\sim$ 0.5 GeV, which corresponds to the oscillation peaks of these experiments. We have plotted $\Delta P_{\mu e}$ with $\delta_{CP}$ varied $\in[-180^\circ,180^\circ]$. Note that the other mixing parameters have been set to the values described in table~\ref{tab:parameters}. The left, middle, and right panels represent the cases of $d_{e\mu}$, $d_{e\tau}$, and $d_{\mu\tau}$, respectively. In the oscillograms, the darker shades of blue contours correspond to increasing value of $\Delta P_{\mu e}$. We list our observations as follows:

\begin{itemize}
    \item  In the presence of $\phi_{e\mu}$ (left panel) and $\phi_{e\tau}$ (middle panel), we observe a similar pattern of degeneracies in the parameter space $\delta_{CP}$ for both the DUNE (top panel) and T2HK (bottom panel) baselines.
    
    \item However, the dependency of $\phi_{\mu\tau}$ (right panel) for both DUNE (top-right panel) and T2HK (bottom-right panel) is significantly different than that of $\phi_{e\mu}$ and $\phi_{e\tau}$. For a constant $\phi_{\mu\tau}$, the variation of $\rm \Delta P_{\mu e}$ with changes in $\delta_{CP}$ is found to be notably small as indicated by the narrow range of the colorbar.
\end{itemize}

The degenerate regions present can directly impact the measurement of the CP violation phase, as discussed in the subsequent sections.
\subsection{Effect on CP asymmetry}
In this work, we focus on examining the potential influence of dark NSI on the CP-measurement capabilities of long-baseline experiments. In neutrino oscillation experiments, the CP-violating effects can be characterized by a quantity known as the CP-asymmetry parameter, which is defined for the appearance channel as,

\begin{equation}\label{def_asym}A^{\mu e}_{CP} = \frac{P_{\mu e}-\bar{P}{\mu e}}{P{\mu e}+\bar{P}{\mu e}},
\end{equation}

\noindent where, $P{\mu e}$ and $\bar{P}_{\mu e}$ are the appearance probabilities of $\nu_e$ and $\bar{\nu_e}$ respectively. The CP asymmetry parameter ($A^{\mu e}_{CP}$) can be an estimate of CP violation as it quantifies the change in oscillation probabilities when the CP phase changes its sign. To obtain the anti-neutrino probability $\bar{P}_{\mu e}$, the signs of all phases (both the standard $\delta_{CP}$ and the dark NSI phases $\phi_{\alpha\beta}$) are reversed, and the dark NSI correction term itself also flips its sign as discussed in section~\ref{sec:formalism}. The shape and magnitude of the CP-asymmetry curve are strongly influenced by the neutrino energy and the baseline traveled.

\begin{figure}[h!]
    \centering
    \hspace{7pt} \includegraphics[width=.7\textwidth]{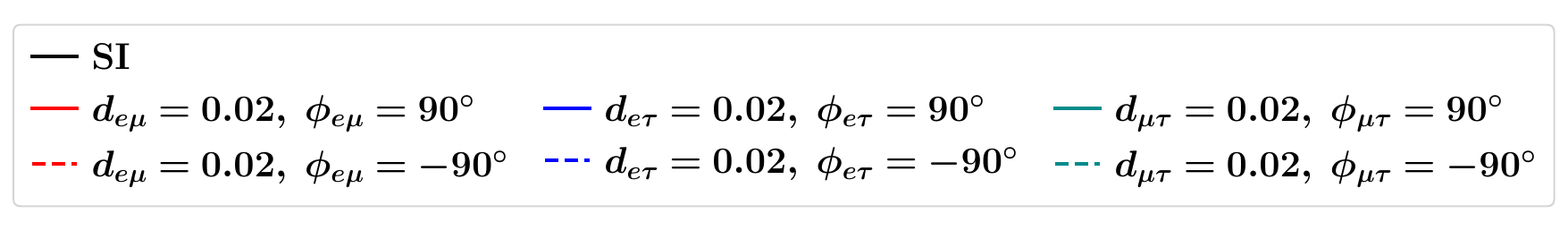}
     \includegraphics[width=.45\textwidth]{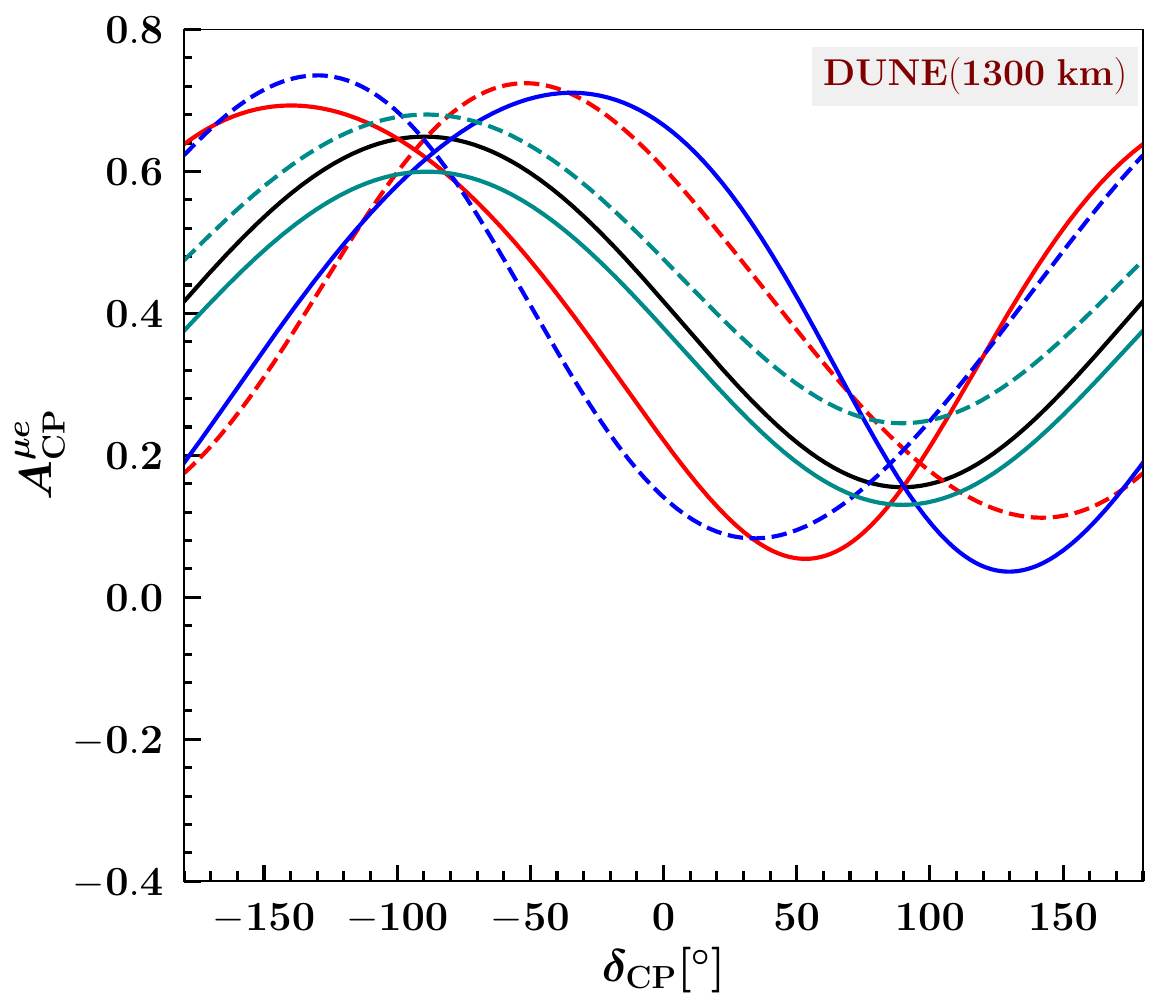}
     \includegraphics[width=.45\textwidth]{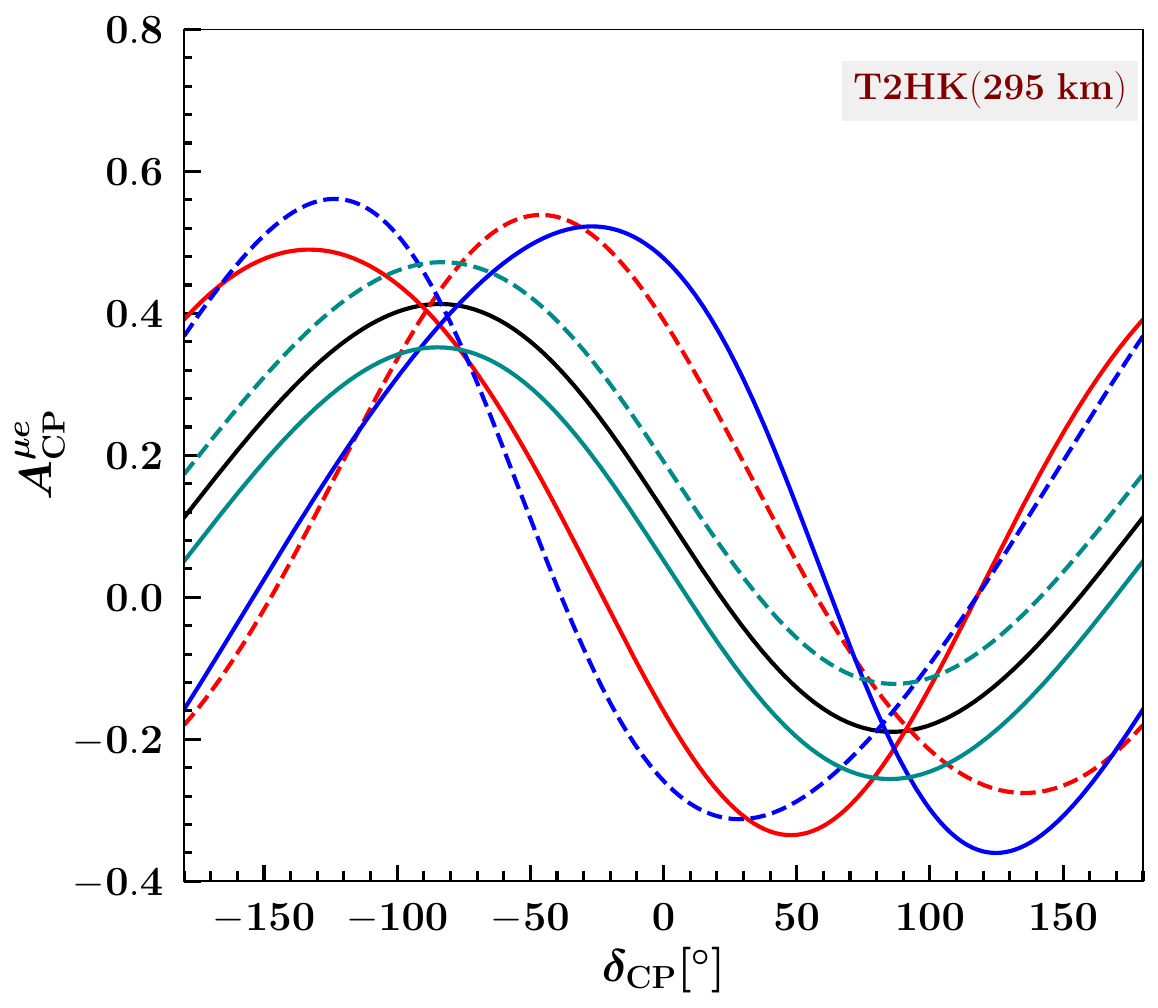}
    \caption{The CP--asymmetry $A^{\mu e}_{CP}$ vs $\delta_{CP}$ plot for DUNE (left--panel) and T2HK (right--panel) in presence of $d_{\alpha\beta}$ at the peak energies. Here, $\theta_{23}$ = 47$^\circ$ and true mass hierarchy = NO. The solid--black curve is for the no dark NSI case, and other coloured solid (dashed) curves are for $\phi_{\alpha\beta} = 90^\circ$ ($\phi_{\alpha\beta} = - 90^\circ$) for chosen $d_{\alpha\beta} = 0.02$. }
    \label{fig:CP_asymmetry}
\end{figure}

 We show $A^{\mu e}_{CP}$ in the presence of dark NSI as a function of $\delta_{CP}$ at the baselines and peak energies of DUNE and T2HK in Figure \ref{fig:CP_asymmetry}. Note that, the peak energies for DUNE have been considered as 2.5 GeV, and the same for T2HK is fixed at 0.5 GeV. The solid--black curve in all the plots represents the no-dark NSI case, $d_{\alpha\beta}$ = 0. The solid (dashed) curves in red, blue, and darkcyan are for $d_{\alpha \beta}= 0.02$ and $\phi_{\alpha \beta}=90^\circ$ ($\phi_{\alpha \beta}=-90^\circ$). The observations from figure \ref{fig:CP_asymmetry} are listed below.

\begin{itemize}
    \item The presence of off-diagonal dark NSI parameters ($d_{e\mu}$, $d_{e\tau}$, and $d_{\mu\tau}$) leads to noticeable deviations in the CP-asymmetry $A^{\mu e}_{\mathrm{CP}}$ compared to the SI case. These deviations are strongly correlated with the associated dark NSI phases $\phi_{\alpha\beta}$, with $\phi_{\alpha\beta} = \pm 90^\circ$ producing distinct shifts in both the magnitude and the shape of the $A^{\mu e}_{\mathrm{CP}}$ curves.

    \item Presence of dark NSI can lead to degeneracies in $A^{\mu e}_{\mathrm{CP}}$ among various values of $d_{\alpha \beta}, \phi_{\alpha \beta}$ and $\delta_{CP}$, which could affect the anticipated CP phase measurement at DUNE and T2HK.

\end{itemize}

\begin{figure}[h!]
	\centering
	\includegraphics[width=0.32\linewidth]{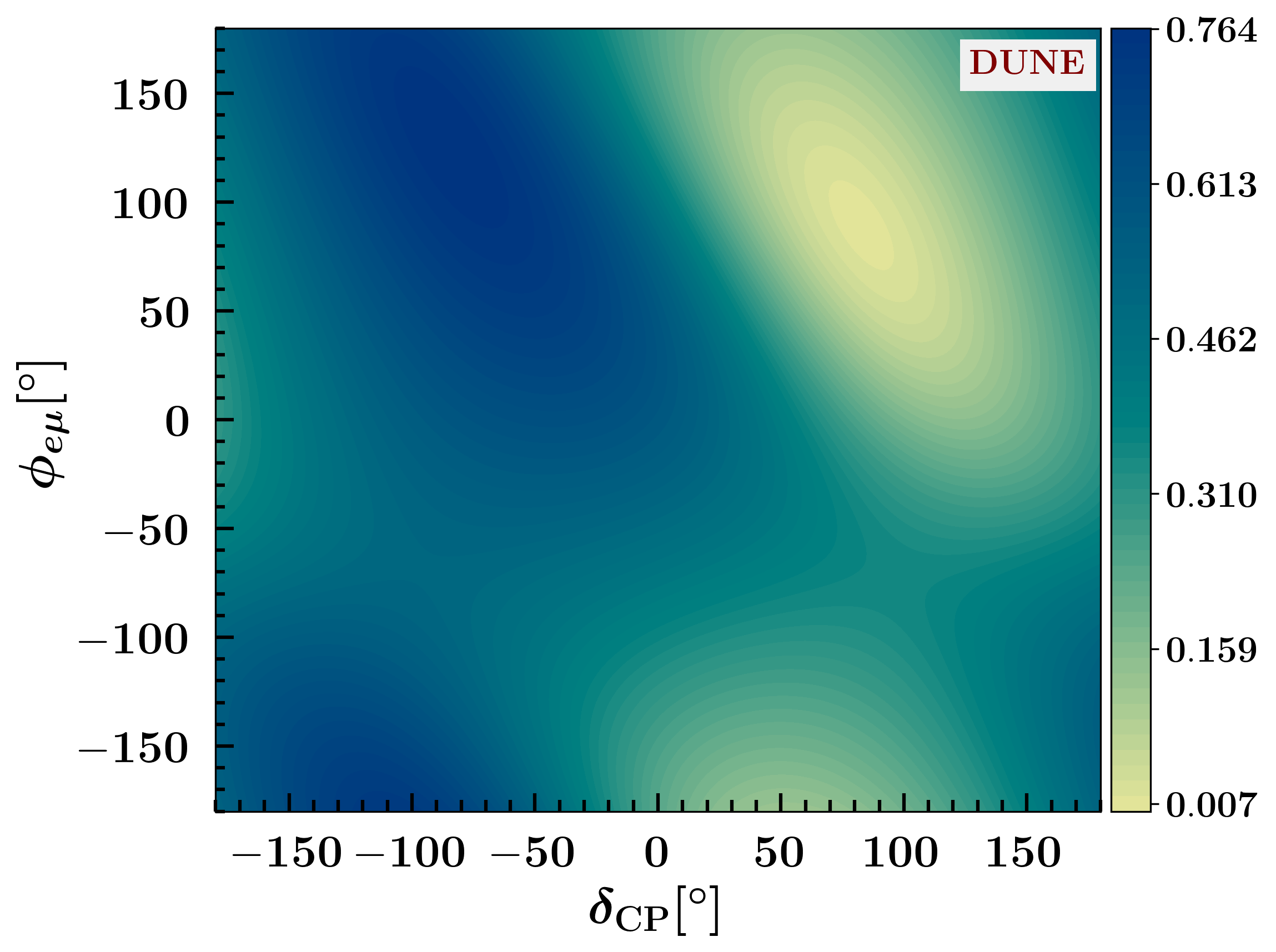}
	\includegraphics[width=0.32\linewidth]{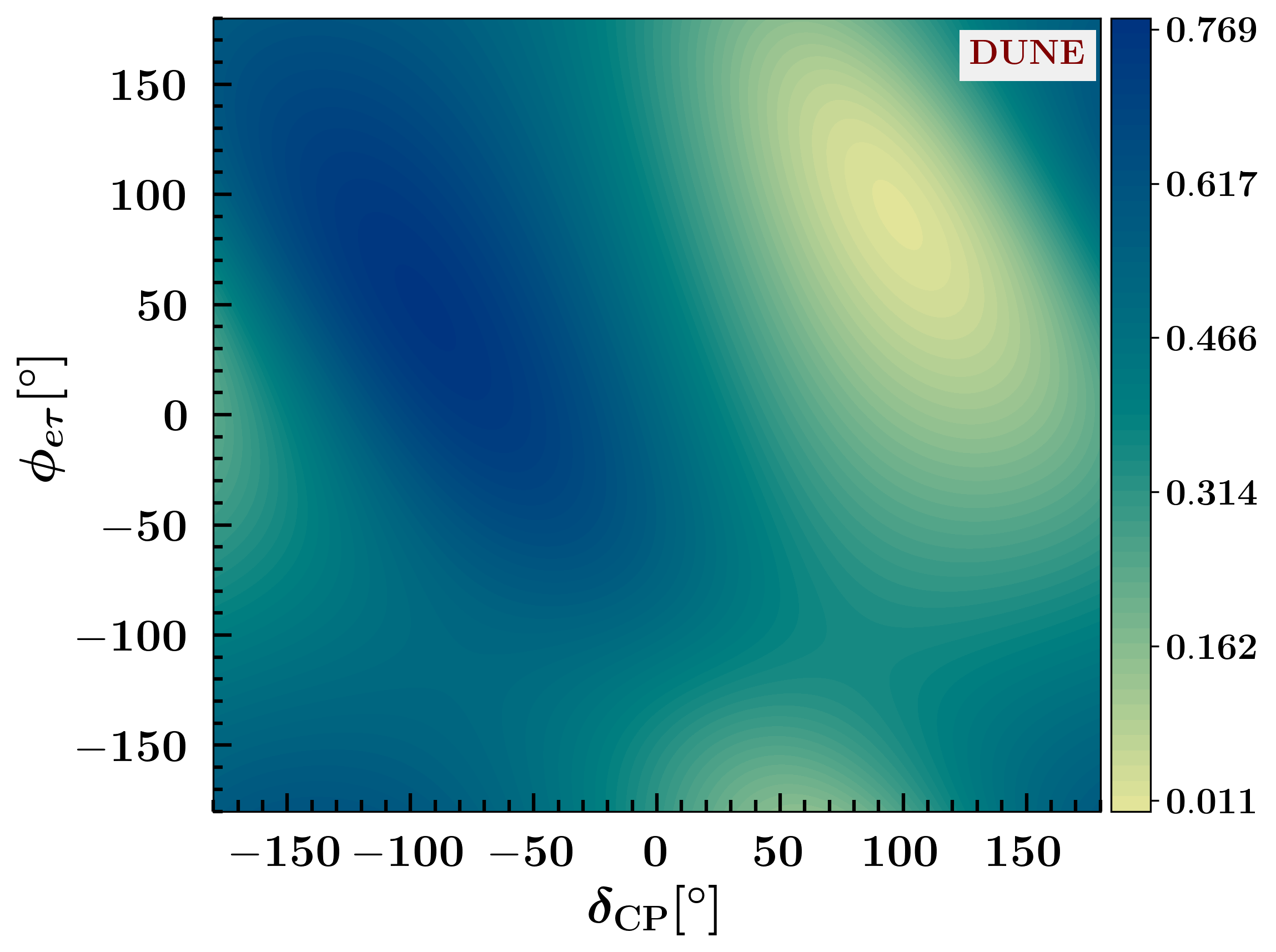}
	\includegraphics[width=0.32\linewidth]{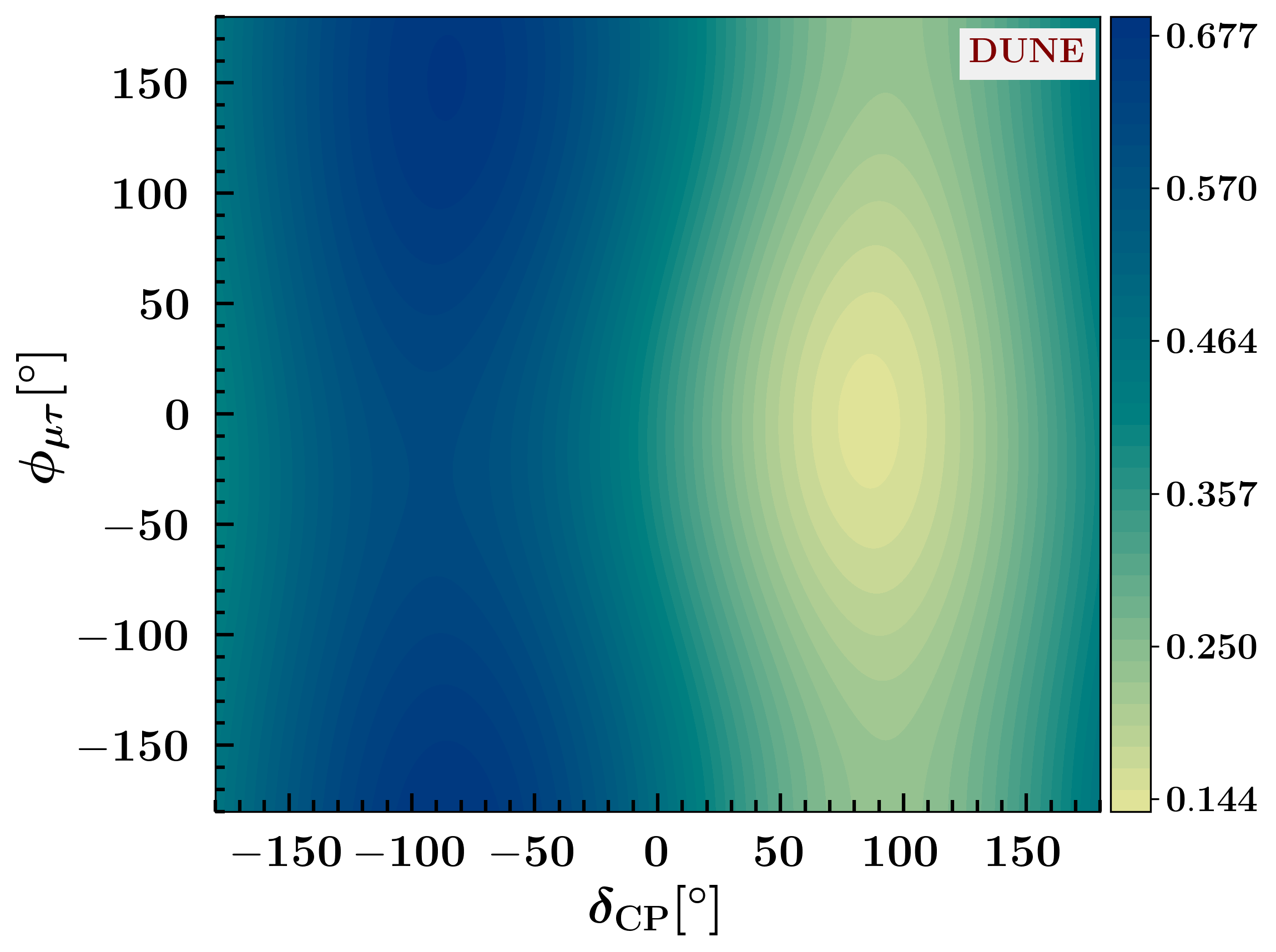}
	\includegraphics[width=0.32\linewidth]{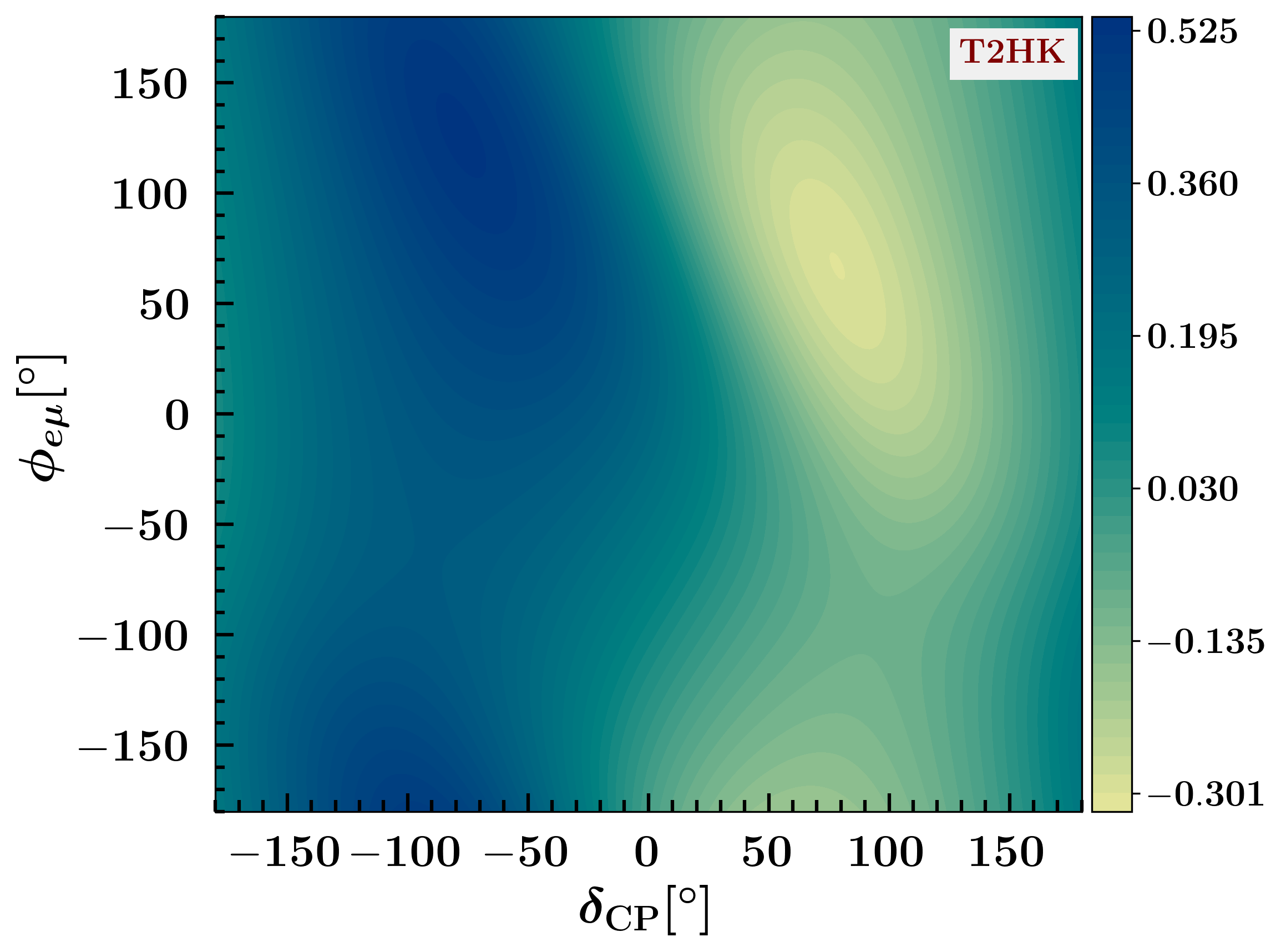}
	\includegraphics[width=0.32\linewidth]{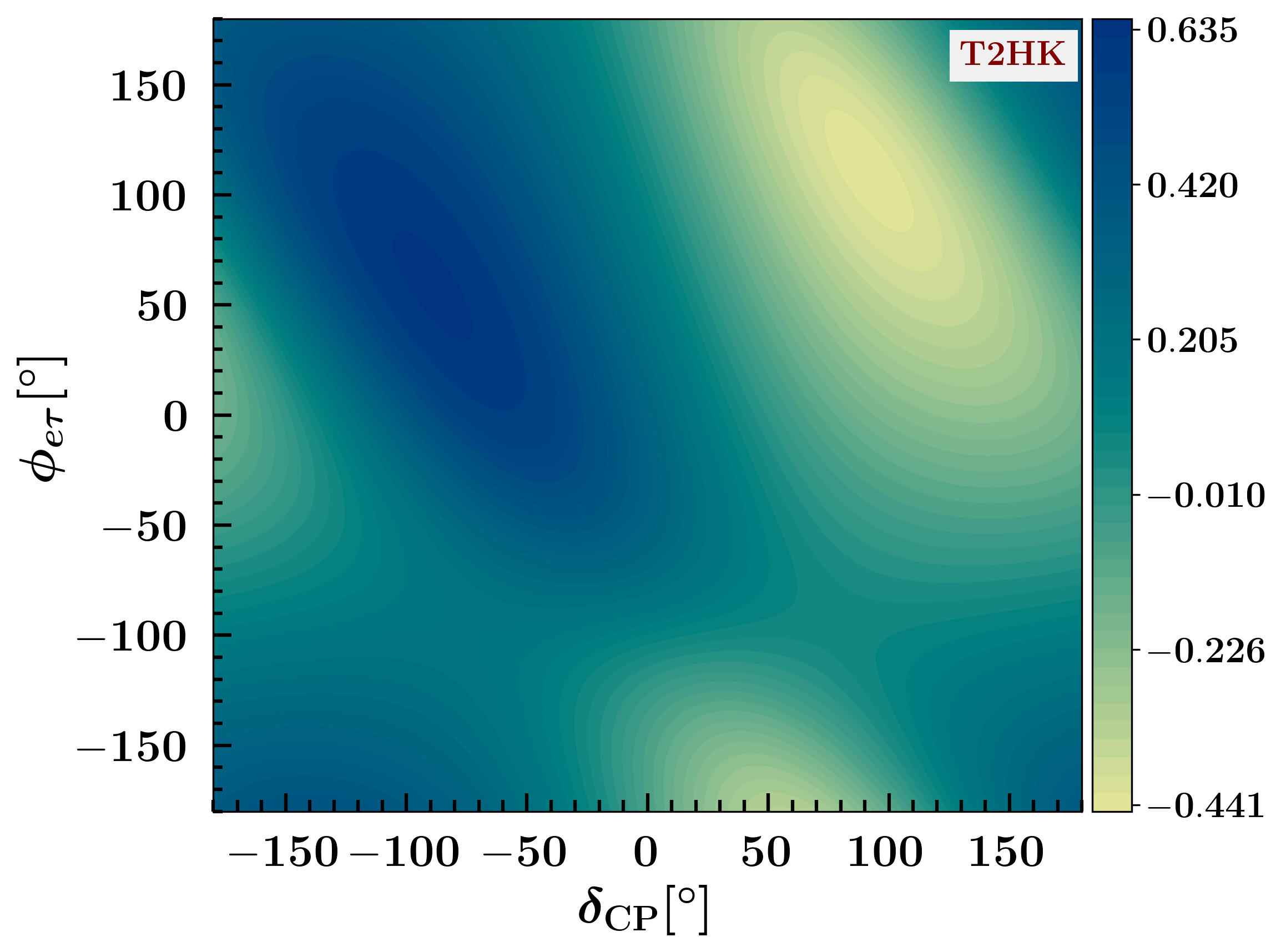}
	\includegraphics[width=0.32\linewidth]{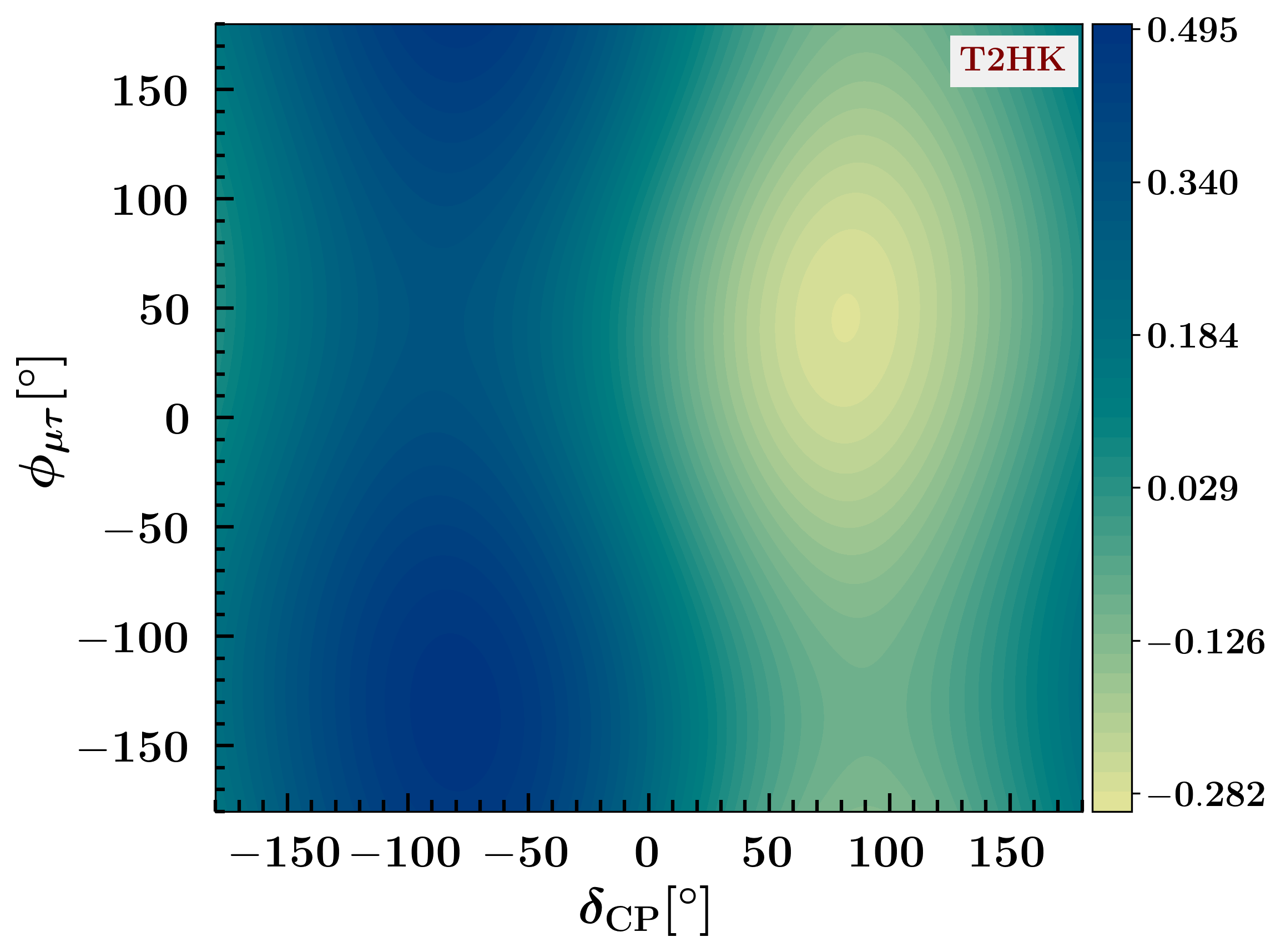}
	\caption{The CP--asymmetry in $\delta_{CP}$-$\phi_{\alpha \beta}$ plane for DUNE (top--panel) and T2HK (bottom--panel) for $d_{\alpha \beta} = 0.2$ at the peak energies. Here, $\theta_{23}$ = 47$^\circ$ and true mass hierarchy = NO. The left panel corresponds to $\phi_{e\mu}$, the middle panel corresponds to $\phi_{e\tau}$, and the right panel corresponds to $\phi_{\mu\tau}$, respectively.}
	\label{fig:CP_asymmetry_2D}
\end{figure}

In the top (bottom) panel of Figure \ref{fig:CP_asymmetry_2D}, we have shown the effect of $\delta_{CP}$ and $\phi_{\alpha \beta}$ on the CP asymmetry parameter for the DUNE (T2HK) baseline. The value of the dark NSI parameters $d_{\alpha \beta}$ is fixed at 0.02. We note that,

\begin{itemize}

    \item For the DUNE baseline (top panbel), the CP asymmetry $A^{\mu e}_{CP}$, shows a strong and complex coupled dependence on both the standard phase $\delta_{CP}$ and the NSI phases $\phi_{e\mu}$ (left panel) and $\phi_{e\tau}$ (middle panel). The maximum (yellow) and minimum (dark blue) values of $A^{\mu e}_{CP}$ shift with the NSI phase rather than occurring at a fixed $\delta_{CP}$ value. This implies that the presence of $d_{e\mu}$ or $d_{e\tau}$ and their associated phases can significantly distort the CP asymmetry measurement and create degeneracies, making it difficult to isolate the true value of $\delta_{CP}$.
    
    \item In contrast, the $A^{\mu e}_{CP}$ is almost completely independent of the NSI phase $\phi_{\mu\tau}$ (right panel). The contours are vertical, demonstrating a clear dependence on $\delta_{CP}$ but negligible variation with $\phi_{\mu\tau}$. The sinusoidal-like dependence of $A^{\mu e}_{CP}$ on $\delta_{CP}$ with maxima near $-90^\circ$ and minima near $+90^\circ$ in the standard case is preserved, regardless of the value of $\phi_{\mu\tau}$.
    
    \item For the T2HK baseline (bottom panel), we observe a similar dependence as that of the DUNE baseline. However, for DUNE baseline (top panel), neutrinos travel through a longer baseline of 1300 km and this matter effect strongly enhances the neutrino probability ($P_{\mu e}$) and suppresses the anti-neutrino probability ($\bar{P}_{\mu e}$), ensuring $P_{\mu e} > \bar{P}_{\mu e}$ regardless of the values of $\delta_{CP}$ or the NSI phases. In contrast, at T2HK (bottom panel) with a baseline of 295 km, matter effects are weak, so the asymmetry mainly comes from genuine CP-phase interference, which can lead to negative $A^{\mu e}_{CP}$ in some regions of the parameter space.
    
\end{itemize}

\subsection{Impact on CP precision measurement sensitivity}

\begin{figure}[h!]
    \centering
     \includegraphics[width=.325\textwidth]{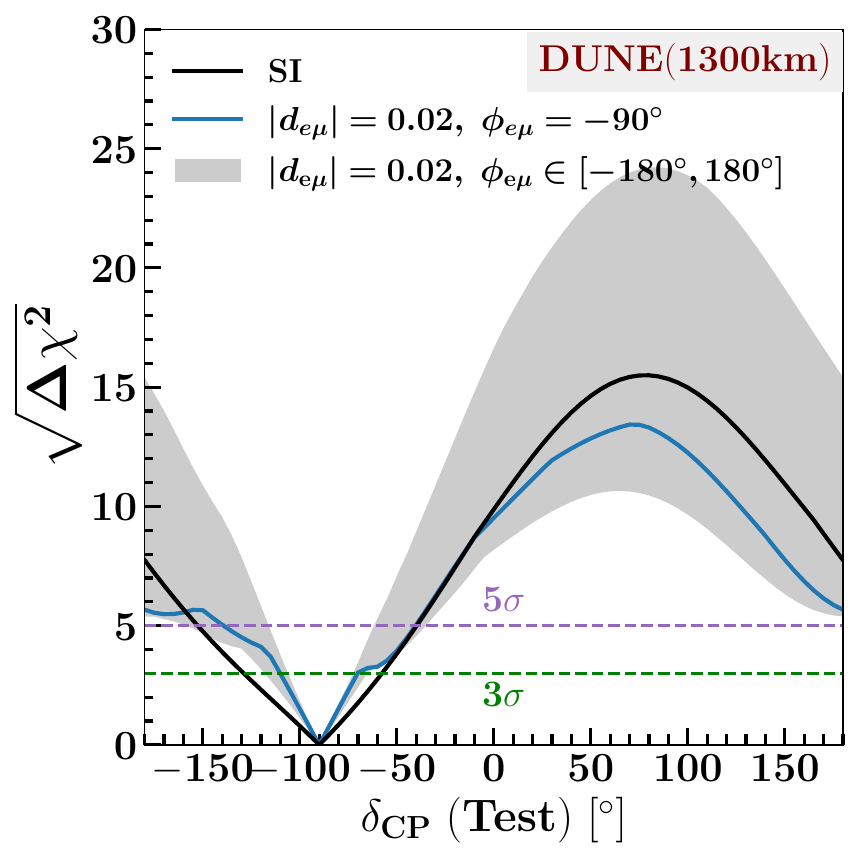}
    \includegraphics[width=.325\textwidth]{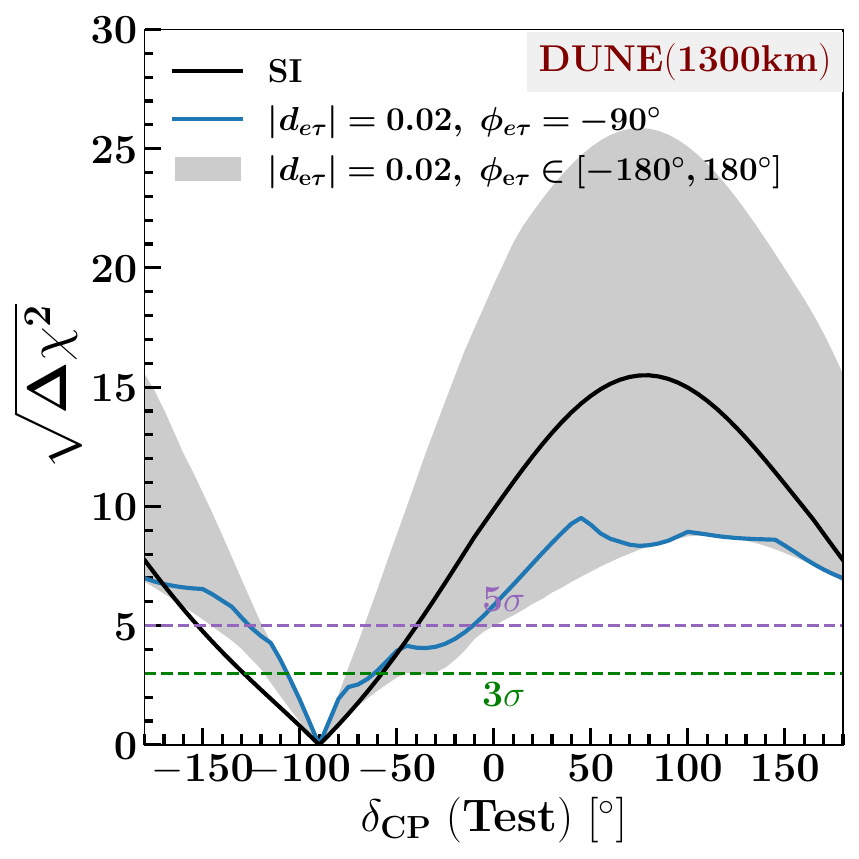}
    \includegraphics[width=.325\textwidth]{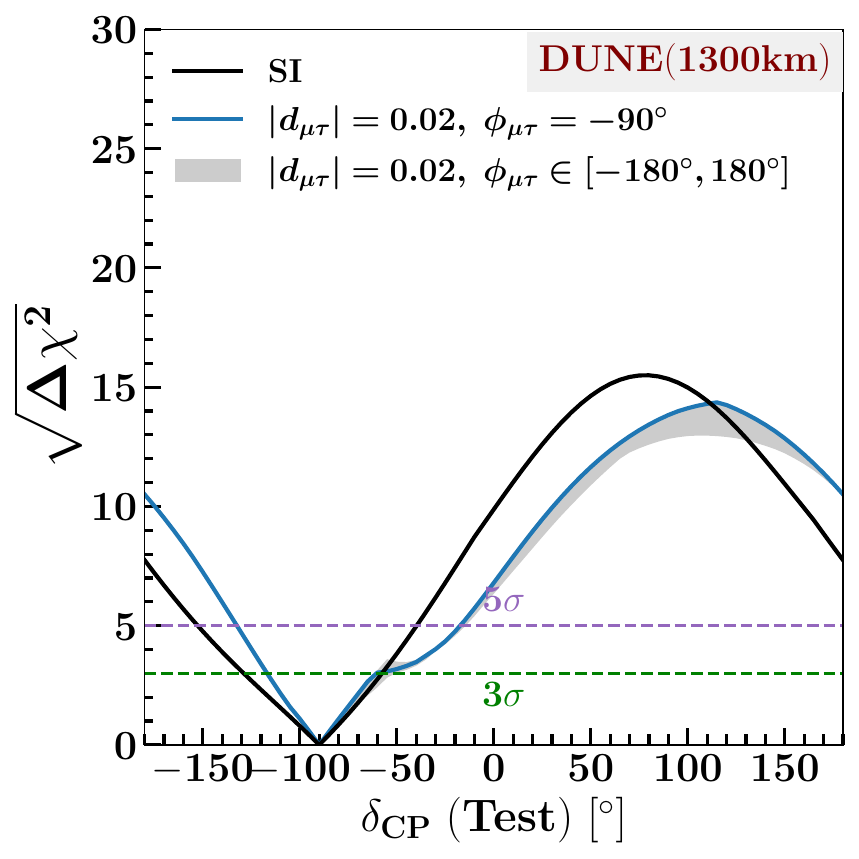}
    \includegraphics[width=.325\textwidth]{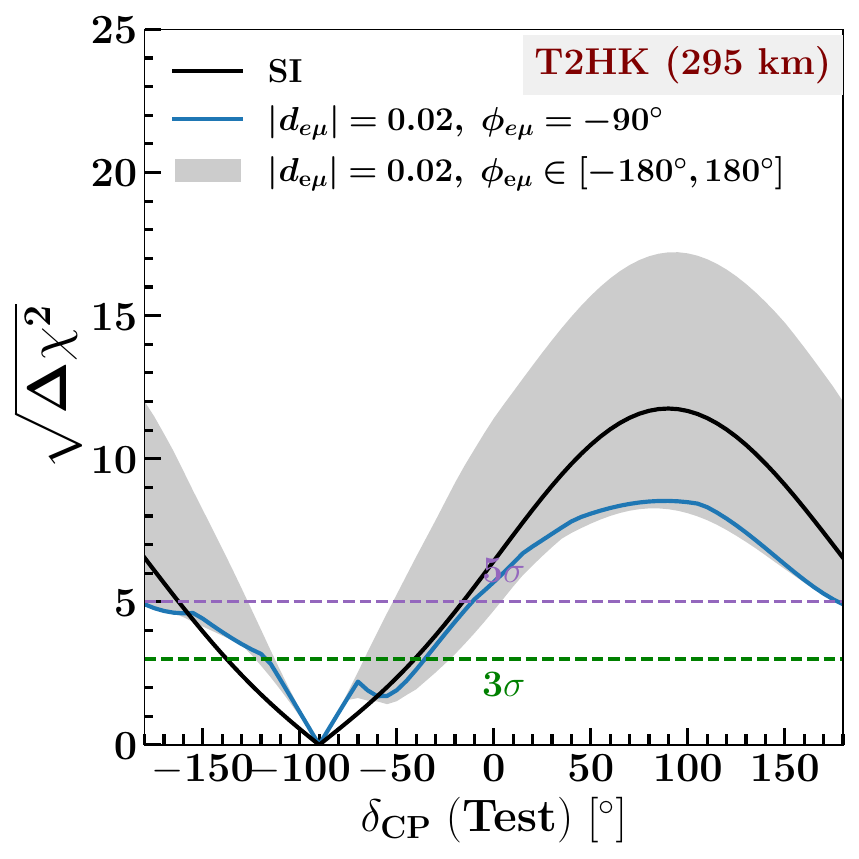}
    \includegraphics[width=.325\textwidth]{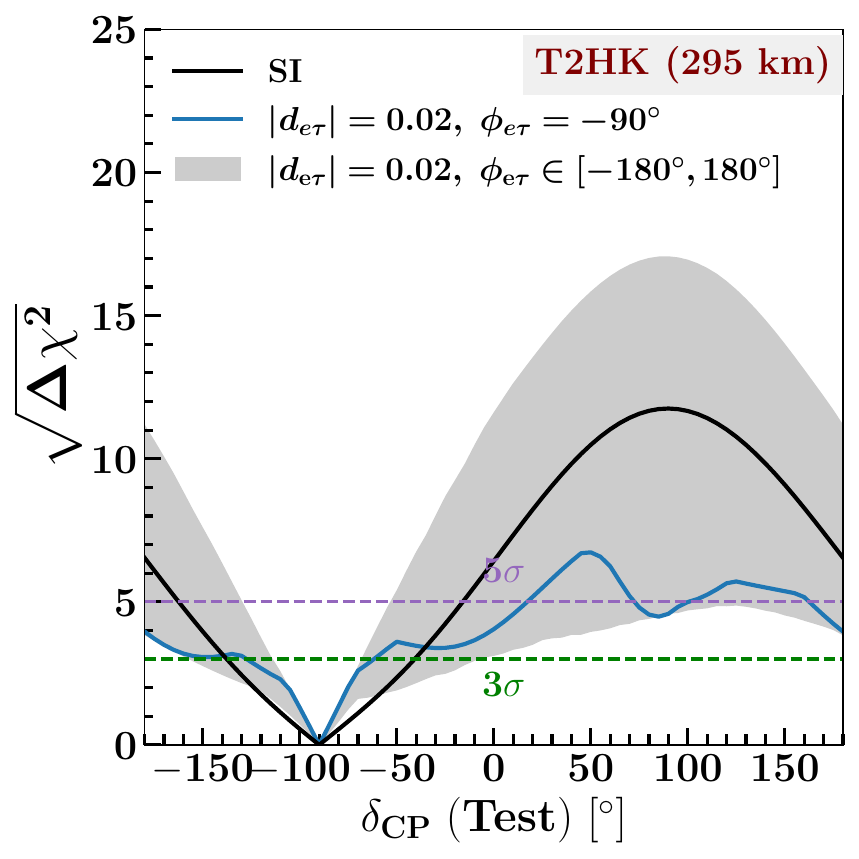}
    \includegraphics[width=.325\textwidth]{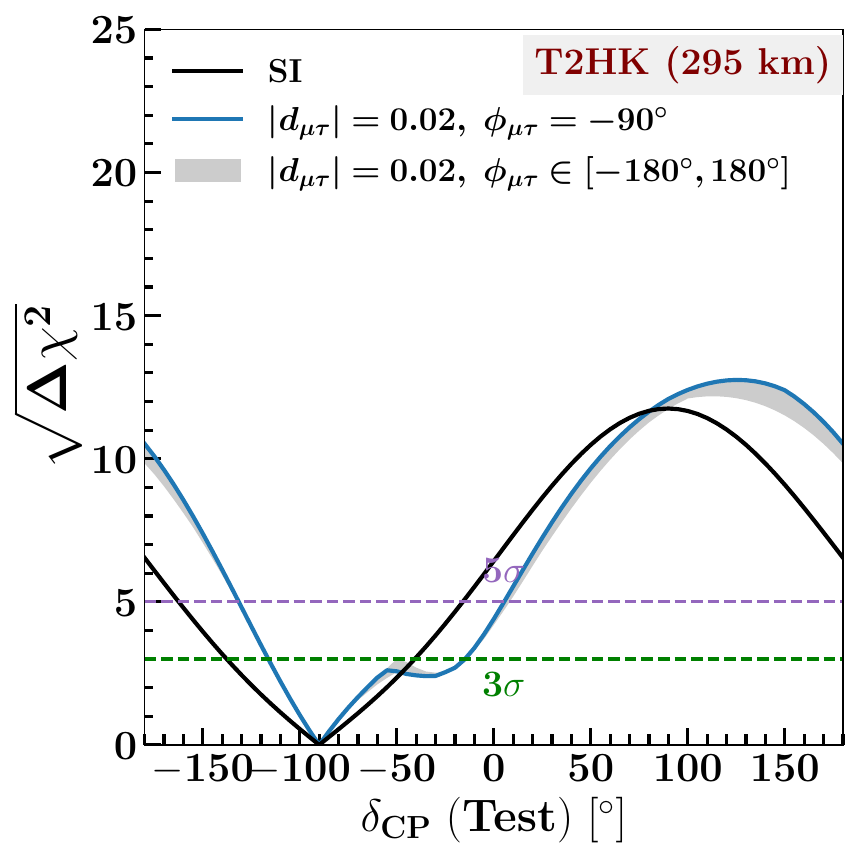}
    \caption{CP precision sensitivity for DUNE (top panel) and T2HK (bottom panel) in presence of off-diagonal dark NSI parameters $d_{e\mu }$ (left), $d_{e \tau}$(middle), and $d_{\mu \tau}$(right). Here, we have considered NO as the true ordering and $\delta_{CP}= -90^\circ$. The gray band corresponds to the true value of $\phi_{\alpha \beta} \in [-180^\circ, 180^\circ]$.}
    \label{fig:CP_prec}
\end{figure}

In figure \ref{fig:CP_prec}, we show the CP phase measurement sensitivity of DUNE (top panel) and T2HK (bottom panel) in the presence of dark NSI elements. We investigate, in the presence of dark NSI, how well an experiment can constrain the values of $\delta_{CP}$ if we assume that the true value of $\delta_{CP}$ is known. We kept the true value of $\delta_{CP}$ fixed at $-90^\circ$ and varied the test $\delta_{CP}$ in the allowed range of $\delta_{CP}~\in [-180^\circ, 180^\circ]$. We have marginalized over the atmospheric mixing angle $\theta_{23}$ and $\Delta m_{31}^2$ in the range mentioned in Table \ref{tab:parameters}. We have also marginalized over the dark NSI parameters in the range $[0, 0.05]$. The black solid line corresponds to the case without dark NSI, and the blue solid line is when $d_{\alpha \beta} = 0.05$ and $\phi_{\alpha \beta} = -90^\circ$. The gray band shows the effect of the phase term $\phi_{\alpha \beta}$ in the entire range $[-180^\circ, 180^\circ]$. We list our observations of this analysis as follows,

\begin{itemize}

	\item For DUNE, in presence of $d_{e\mu}$ (left panel) and $d_{e \tau}$ (middle panel), when the phase $\phi_{\alpha\beta}$ is set to $-90^\circ$, the constraining capability slightly increases as compared to the SI case in the negative half plane of test $\delta_{CP}$. Whereas, we observe a noticeable suppression of sensitivity for the positive $\delta_{CP}$ plane. This can be understood from the behavior of the CP-asymmetry parameter $A_{CP}$ in figure~\ref{fig:CP_asymmetry} and \ref{fig:CP_asymmetry_2D}. In the positive $\delta_{CP}$ plane, we observe a large $A_{CP}$ which varies sharply with $\delta_{CP}$ aiding to the enhancement in sensitivity in this region. The $\delta_{CP}-\phi_{\alpha\beta}$ degeneracy, implied by the flatter contours in figure \ref{fig:CP_asymmetry_2D}, can be responsible for the broad uncertainty bands and the reduced sensitivity shown in the left and middle panel of figure~\ref{fig:CP_prec}.
	
	\item For $d_{\mu \tau}$ (right panel), the dark NSI phase $\phi_{\mu \tau}$ has nominal effect on the CP-precision sensitivity. It is expected from the behaviour of $A_{CP}$ in figure \ref{fig:CP_asymmetry_2D}, where it has very little dependence on $\delta_{CP}$. Moreover, the enhancement (suppression) in CP-precession sensitivity on the positive (negative) $\delta_{CP}$ half plane when the phase $\phi_{\mu\tau}=-90^\circ$ is also expected from the behaviour of $A_{CP}$.
	
	 \item In case of T2HK, the behavior of the CP-precision sensitivity is similar to that of DUNE. However, we observe milder distortions and narrower gray bands in the CP-precision plots as a result of its shorter baseline and reduced matter effects.

\end{itemize}

\subsection{Impact on CP violation sensitivity}

\begin{figure}[h!]
    \centering
     \includegraphics[width=.325\textwidth]{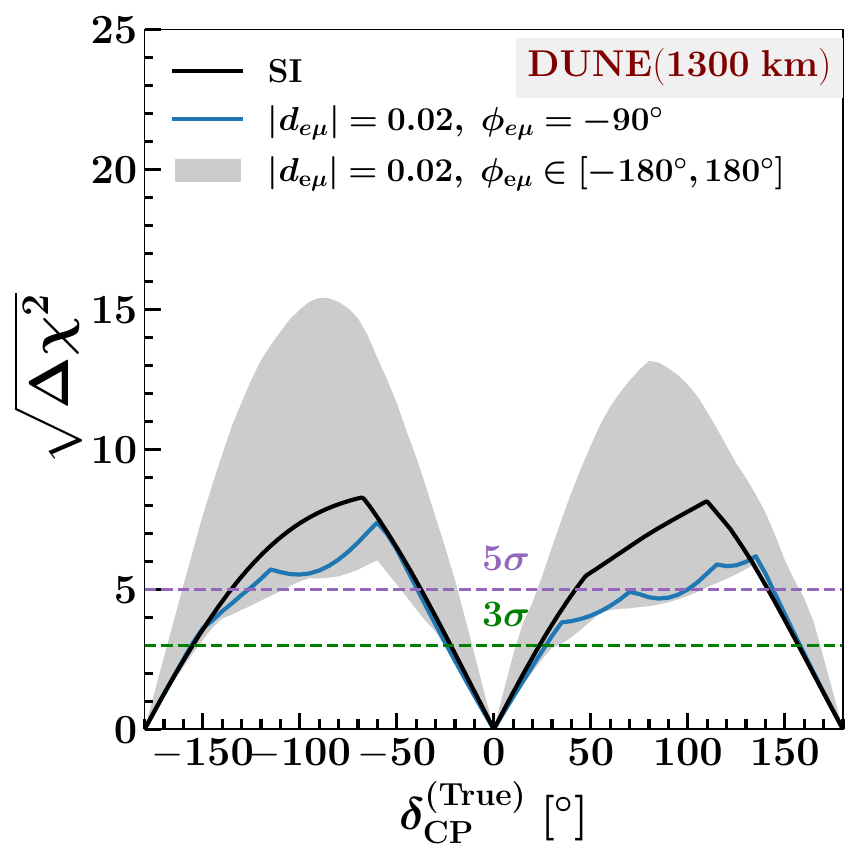}
    \includegraphics[width=.325\textwidth]{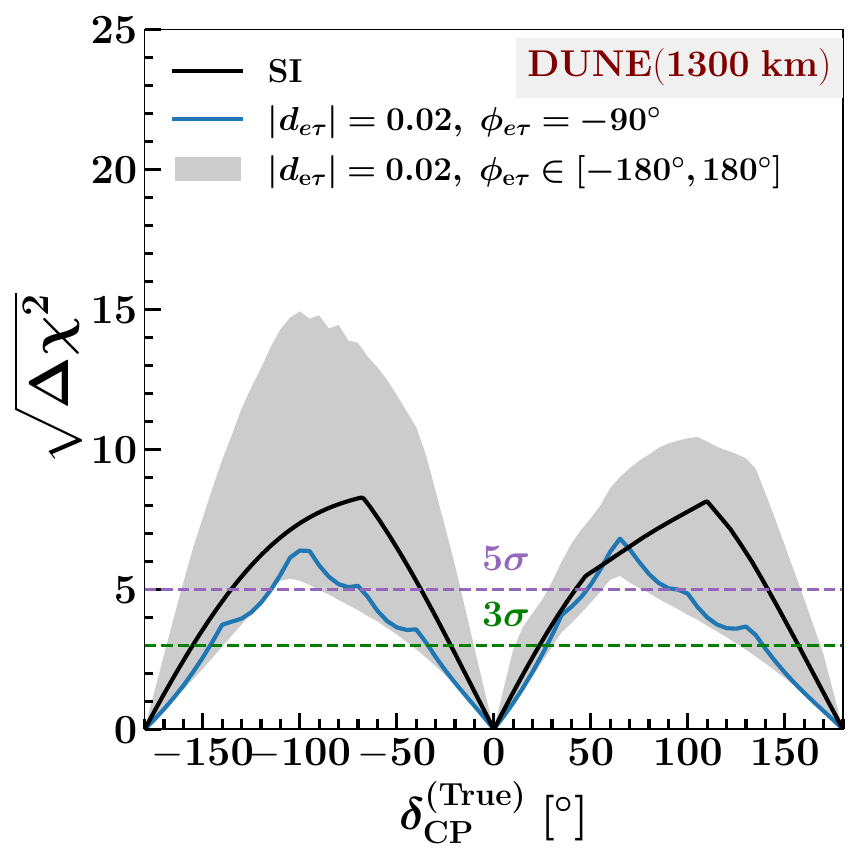}
    \includegraphics[width=.325\textwidth]{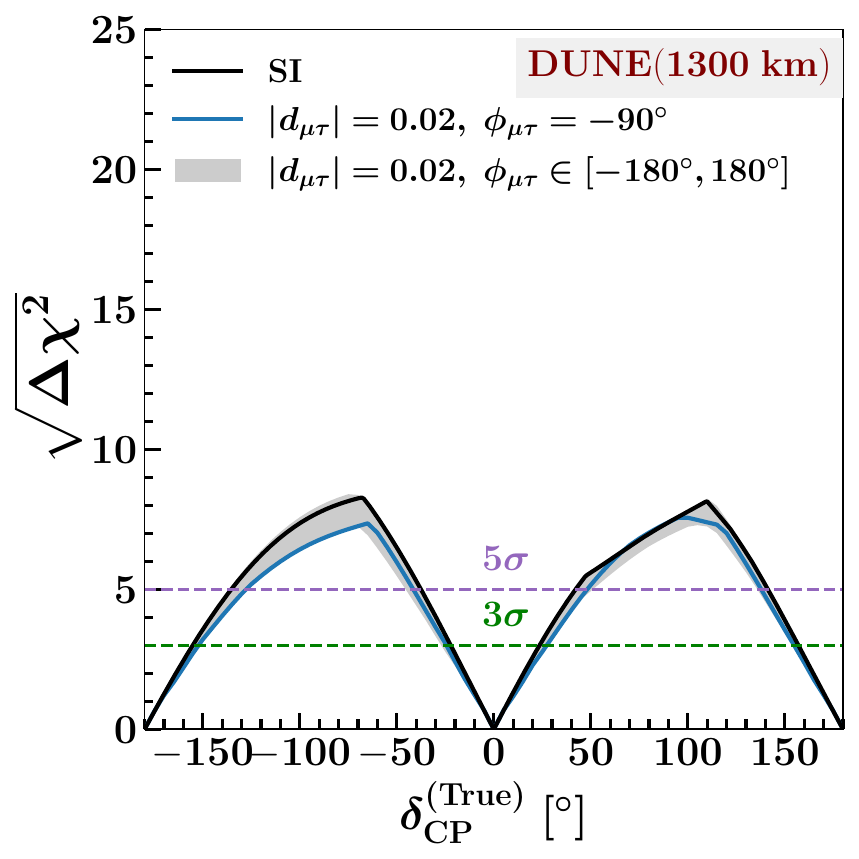}
    \includegraphics[width=.325\textwidth]{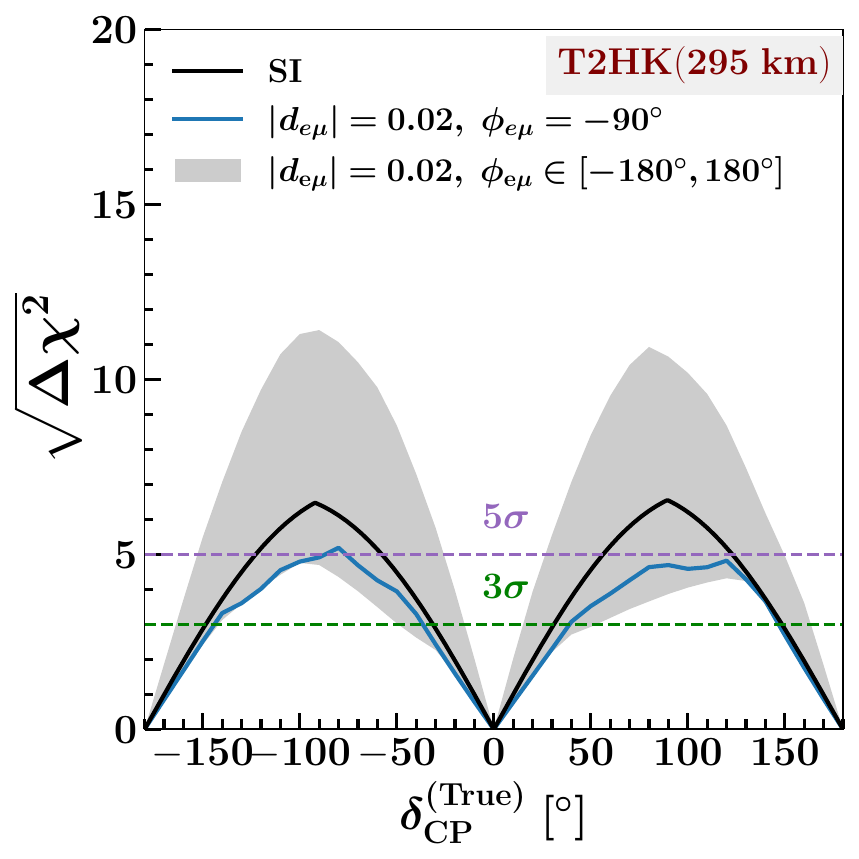}
    \includegraphics[width=.325\textwidth]{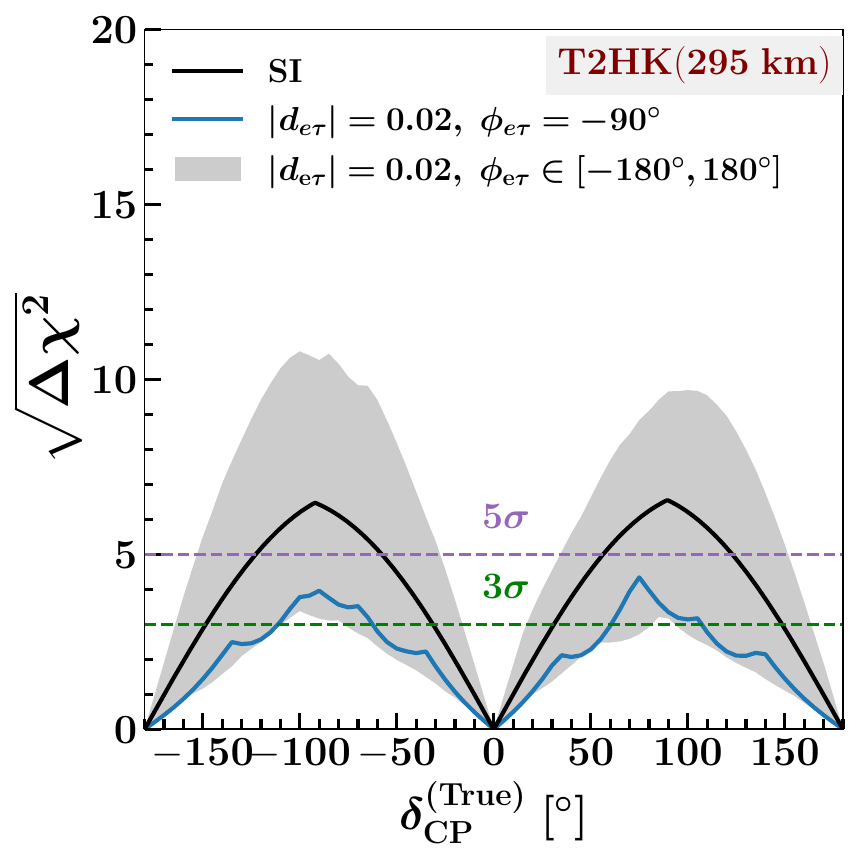}
    \includegraphics[width=.325\textwidth]{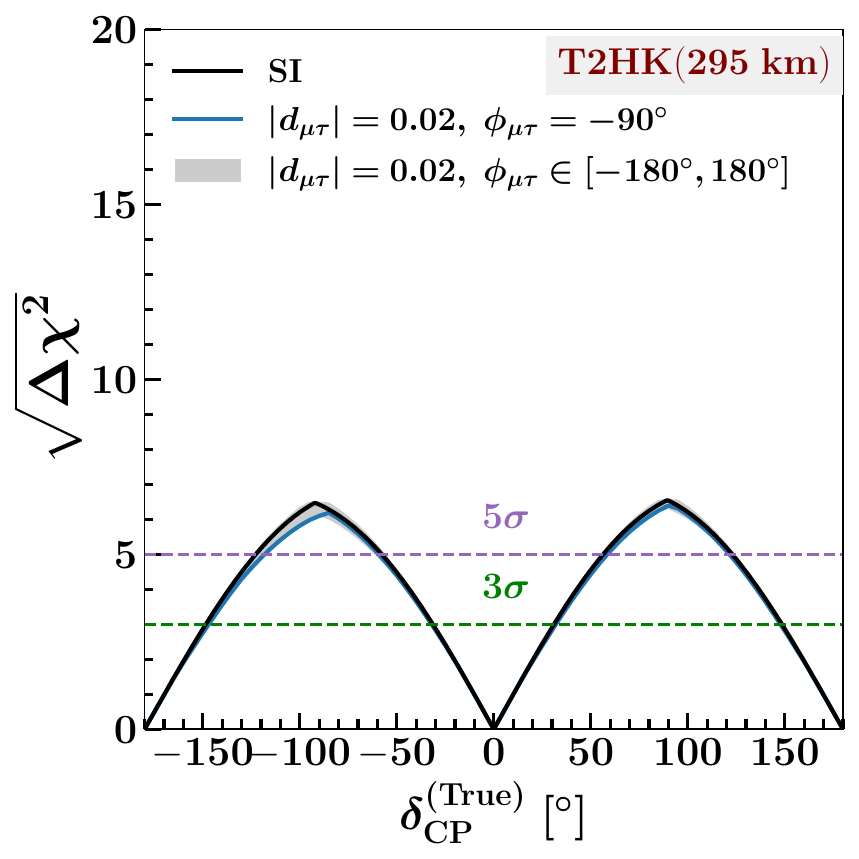}
    \includegraphics[width=.325\textwidth]{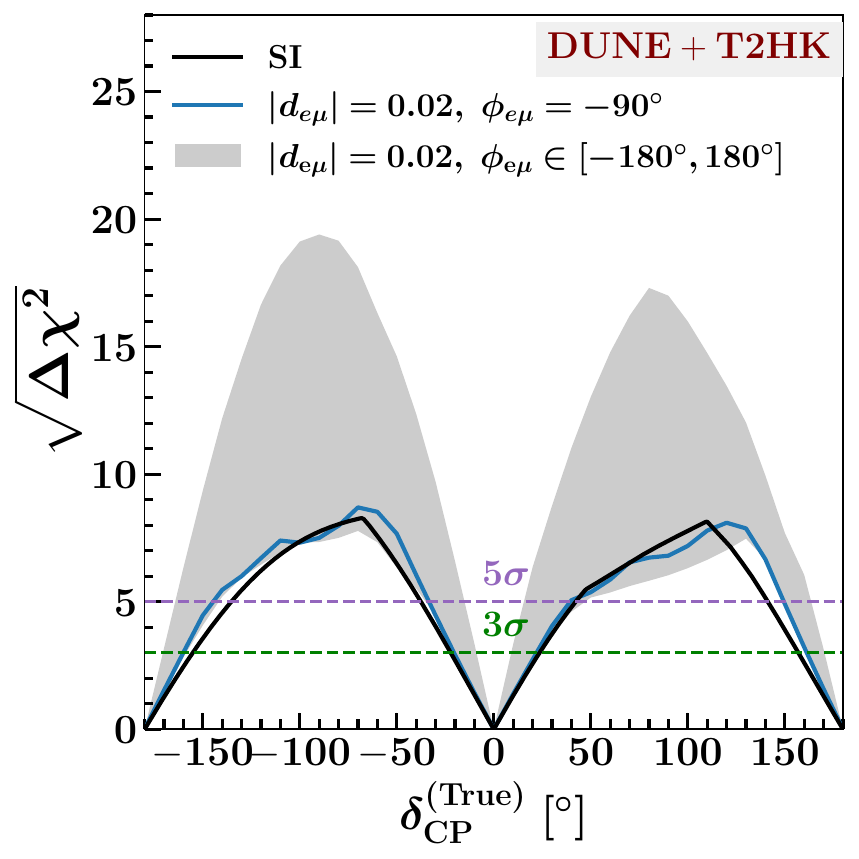}
    \includegraphics[width=.325\textwidth]{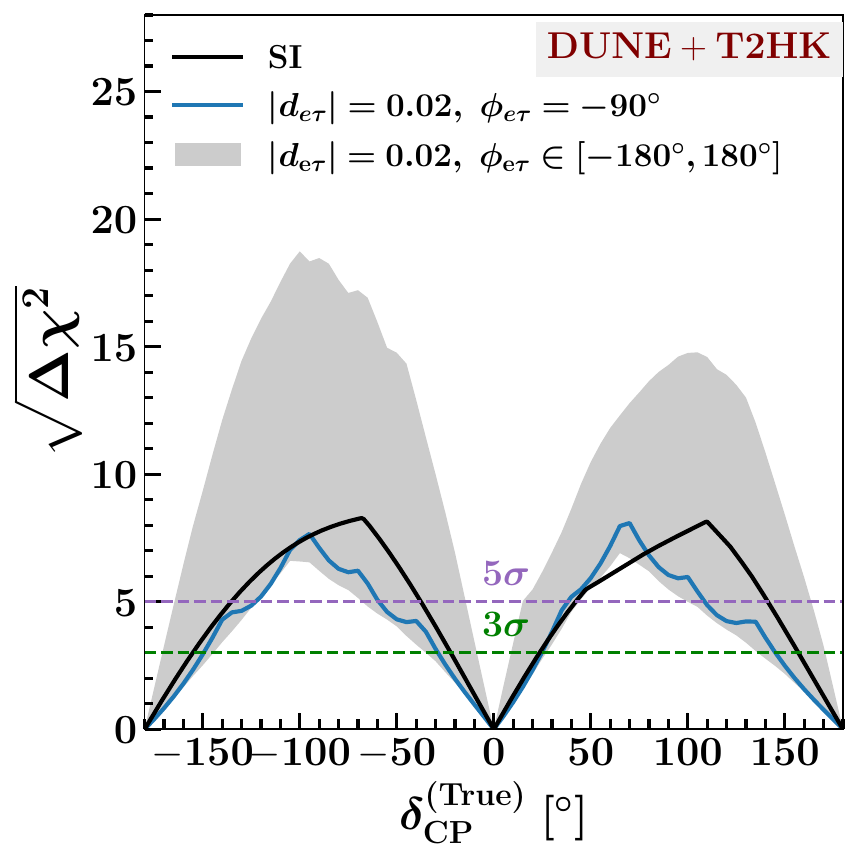}
    \includegraphics[width=.325\textwidth]{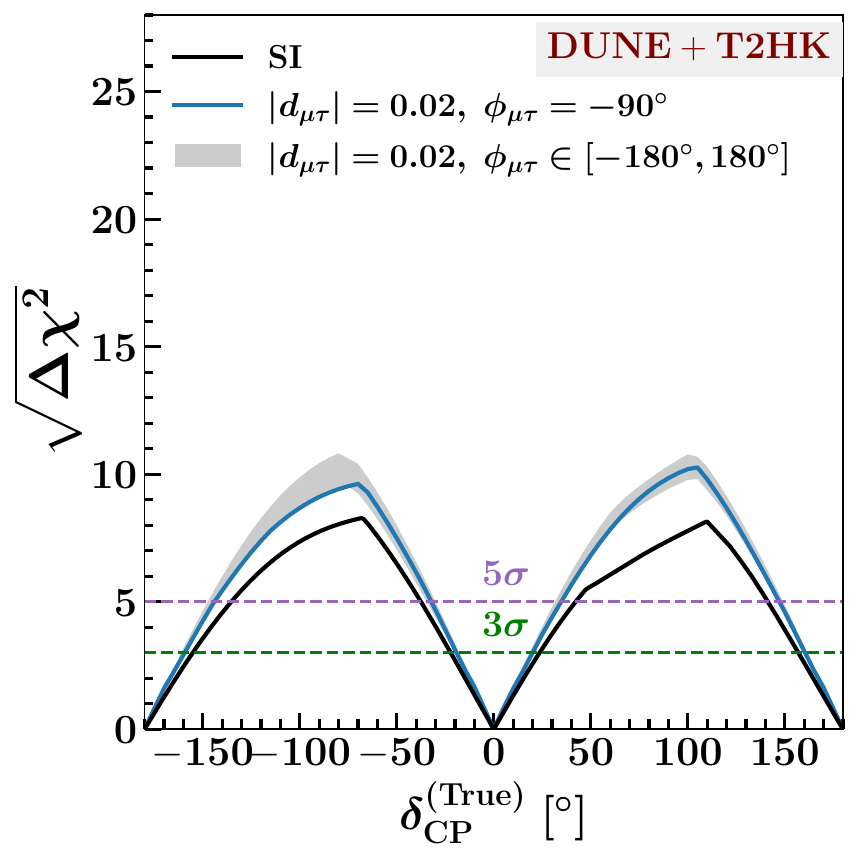}
    \caption{CPV sensitivity of DUNE (top panel), T2HK (middle panel), and DUNE+T2HK (bottom panel) in presence of off-diagonal dark NSI parameters $d_{e\mu }$ (left), $d_{e \tau}$(middle), and $d_{\mu \tau}$(right). Here NO is the true ordering and the value of $\delta_{CP}= -90^\circ$. The gray band corresponds to the true value of $\phi_{\alpha \beta} \in [-180^\circ, 180^\circ]$.}
    \label{fig:CPV}
\end{figure}

In figure \ref{fig:CPV}, we show the impact of dark NSI elements on the CP violation sensitivity at DUNE. We define the CPV sensitivity as, 
\begin{equation}
{\Delta \chi}^{2}_{\rm CPV}~(\delta^{\rm true}_{\rm CP}) = {\rm min}~\left[\chi^2~(\delta^\text{true}_{CP},\delta^\text{test}_{CP}=0),~\chi^2 (\delta^\text{true}_{CP},\delta^\text{test}_{CP}=\pm \pi)\right ].
\end{equation}
 We have obtained the sensitivities by varying the true values of $\delta_{CP}$ in the allowed range [-$\pi$, $\pi$]. The true values of other mixing parameters used in this analysis are listed in table~\ref{tab:parameters}. In the test spectrum of $\delta_{CP}$, we have only considered the CP-conserving values, 0 and $\pm$ $\pi$. We have marginalized $\theta_{23}$ and $\Delta m_{31}^2$ over the allowed 3$\sigma$ ranges~\cite{Esteban:2024eli}. Additionally, we have also marginalized over the dark NSI parameters $\eta_{\alpha \beta}$ in range 0 to 0.05 and have minimized the $\chi^2$ over all the marginalization ranges. In the plots, we have shown the effect of the phase $\phi_{\alpha \beta}$, with the gray band. The blue solid line corresponds to the case of $\phi_{\alpha \beta} = -90^\circ$. The green (violet) dashed lines are drawn as a reference, which corresponds to 3$\sigma$ (5$\sigma$) confidence level (CL). The top panel corresponds to DUNE, the middle panel to T2HK, and the right panel corresponds to the synergy between DUNE and T2HK. The observations from the figure are listed below.

 \begin{itemize}
     
     \item The presence of dark NSI can significantly impact the CPV sensitivity of DUNE. For the chosen case of $d_{e\mu}=0.02$ (left panel), depending on the value of $\phi_{e\mu}$, the sensitivity may be enhanced or suppressed, as can be seen from the gray band. However, when $\phi_{e\mu} = -90^\circ$, the sensitivity is suppressed as compared to the SI case. We observe a similar behavior for $d_{e \tau}$ (middle panel). In the presence of $d_{\mu \tau}$ (right panel), the effect of $\phi_{\mu\tau}$ is nominal.

     \item For T2HK, the sensitivity follows a similar trend to that of DUNE. Though the sensitivity of DUNE is higher than that of T2HK as the matter effect for DUNE will be higher than that of T2HK due to its longer baseline.

     \item When we combine the data from DUNE and T2HK, the sensitivity is enhanced as compared to the individual cases. This is because of the larger statistics obtained from a wider spectrum of energies. A significant enhancement in the sensitivity is observed for all three cases by performing a joint analysis. In the presence of $d_{e \mu}$ and $d_{e \tau}$, the $\phi_{\alpha \beta} = -90^\circ$ case converges to the SI case within the combined framework. Interestingly, for $d_{\mu \tau}$, the sensitivity lies above the standard sensitivity for all true values of $\delta_{CP}$ under this synergy.
     
 \end{itemize}

\section{Summary and concluding remarks}\label{sec:summary}

In this paper, we perform a comprehensive study of neutrino interactions with complex scalar field DM, referred to as dark NSI. Dark NSI introduces an energy-dependent correction to the neutrino Hamiltonian which can be considered as a medium-dependent correction to the mass-squared term. It should be noted that the dark NSI correction for neutrinos and antineutrinos has opposing signs, giving rise to non-trivial phenomenological implications, particularly in the measurement of the leptonic CP-violating phase $\delta_{CP}$. Neutrino oscillation experiments, owing to their sensitivity to coherent forward scattering, provide an ideal framework to probe these effects. We first investigate the impact of both diagonal and off-diagonal dark NSI parameters on neutrino oscillation probabilities for two representative baselines, 295 km and 1300 km, corresponding to the long-baseline experiments T2HK and DUNE, respectively. The moduli $d_{\alpha \beta}$ and the phases $\phi_{\alpha \beta}$ introduced in the presence of dark NSI can significantly affect the CP phase measurement in long baseline experiments. We subsequently examine the CP violation sensitivity and the CP-precision measurement of DUNE and T2HK in the presence of off-diagonal dark NSI elements and their corresponding phases.

Our results demonstrate that off-diagonal dark NSI elements can have a pronounced impact on CP sensitivity of both the experiments. As we do not know the value of the phase $\phi_{\alpha \beta}$, it can either enhance or suppress the sensitivity. In particular, when $|d_{e \mu}|~=~|d_{e \tau}|~=~0.02$, and the corresponding true value of $\phi_{\alpha \beta}~=-90^\circ$, the sensitivity is substantially suppressed for both DUNE and T2HK. However, a synergy of the two experiments is able to restore the sensitivity to a level comparable to that obtained in the standard oscillation scenario without dark NSI. In contrast, the $|d_{\mu \tau}|$ element and its associated phase $\phi_{\mu \tau}$ have only a marginal impact on the CP sensitivity of DUNE and T2HK. For $|d_{\mu \tau}|~=~0.02$, the sensitivity remains below that of the standard interaction case for all values of $\phi_{\mu \tau}$. Nevertheless, combining data of DUNE and T2HK can substantially enhance the CP sensitivity, lifting it above the standard interaction case for all values of $\phi_{\alpha \beta}$. These results highlight the importance of synergy in disentangling dark NSI effects and improving the robustness of CP phase measurements in future high-precision neutrino oscillation experiments.

\section*{Acknowledgments}
  DB would like to thank DST INSPIRE Fellowship (DST/INSPIRE Fellowship/2022/IF220161) for providing financial support to carry out this work. The authors acknowledge the DST SERB grant CRG/2021/002961. AM acknowledges the support of DST SERB grant  xPHYSPNxDST00627xxBB006-ZBSA-3237 titled ``Indian Institutions-Fermilab Collaboration in Neutrino Physics'' for financial support. The authors would like to thank Dr. Manoranjan Dutta for fruitful discussions at the early stage of this work.

\bibliographystyle{JHEP}
\bibliography{main}

\end{document}